\documentclass[a4paper,clock]{article}
\usepackage[dvips]{epsfig}
\usepackage[dvips]{graphics}
\usepackage[dvips]{color}
\usepackage{amssymb}

\catcode`\@=11
\def\marginnote#1{}
\newcount\hour
\newcount\minute
\newtoks\amorpm
\hour=\time\divide\hour by60 \minute=\time{\multiply\hour by60
\global\advance\minute by-\hour}\edef\standardtime{{\ifnum\hour<12
\global\amorpm={am}%
        \else\global\amorpm={pm}\advance\hour by-12 \fi
        \ifnum\hour=0 \hour=12 \fi
        \number\hour:\ifnum\minute<10
0\fi\number\minute\the\amorpm}}
\edef\militarytime{\number\hour:\ifnum\minute<10
0\fi\number\minute}

\def\draftlabel#1{{\@bsphack\if@filesw {\let\thepage\relax
   \xdef\@gtempa{\write\@auxout{\string
      \newlabel{#1}{{\@currentlabel}{\thepage}}}}}\@gtempa
   \if@nobreak \ifvmode\nobreak\fi\fi\fi\@esphack}
        \gdef\@eqnlabel{#1}}
\def\@eqnlabel{}
\def\@vacuum{}
\def\draftmarginnote#1{\marginpar{\raggedright\scriptsize\tt#1}}
\def\draft{\oddsidemargin -.5truein
        \def\@oddfoot{\sl preliminary draft \hfil
        \rm\thepage\hfil\sl\today\quad\militarytime}
        \let\@evenfoot\@oddfoot \overfullrule 3pt
        \let\label=\draftlabel
        \let\marginnote=\draftmarginnote

\def\@eqnnum{(\theequation)\rlap{\kern\marginparsep\tt\@eqnlabel}%
\global\let\@eqnlabel\@vacuum}  }

\def\numberbysection{\@addtoreset{equation}{section}
        \def\theequation{\thesection.\arabic{equation}}}

\def\underline#1{\relax\ifmmode\@@underline#1\else
 $\@@underline{\hbox{#1}}$\relax\fi}

\catcode`@=12 \relax

\topmargin by -\headsep

\def\br{\begin{eqnarray}}
\def\er{\end{eqnarray}}
\def\be{\begin{equation}}
\def\ee{\end{equation}}

\def\({\left(}
\def\){\right)}

\relax

\def\a{\alpha}

\def\b{\beta}

\def\d{\delta}

\def\g{\gamma}

\def\l{\lambda}
\def\L{\Lambda}

\def\pa{\partial}

\def\ra{\rightarrow}

\def\th{\theta}

\def\tp0{\Theta_{+}^{(0)}}
\def\tm0{\Theta_{-}^{(0)}}

\def\vp{\varphi}

\def\cgh{{\hat {\cal G}}}

\def\f#1#2#3 {f^{#1#2}_{#3}}

\def\win1{{\sf w_{1+\infty}}}

\def\Win1{{\sf W_{1+\infty}}}

\def\rlx{\relax\leavevmode}
\def\inbar{\vrule height1.5ex width.4pt depth0pt}
\def\IZ{\rlx\hbox{\sf Z\kern-.4em Z}}
\def\IR{\rlx\hbox{\rm I\kern-.18em R}}
\def\IC{\rlx\hbox{\,$\inbar\kern-.3em{\rm C}$}}
\def\IN{\rlx\hbox{\rm I\kern-.18em N}}
\def\IO{\rlx\hbox{\,$\inbar\kern-.3em{\rm O}$}}
\def\IP{\rlx\hbox{\rm I\kern-.18em P}}
\def\IQ{\rlx\hbox{\,$\inbar\kern-.3em{\rm Q}$}}
\def\IF{\rlx\hbox{\rm I\kern-.18em F}}
\def\IG{\rlx\hbox{\,$\inbar\kern-.3em{\rm G}$}}
\def\IH{\rlx\hbox{\rm I\kern-.18em H}}
\def\II{\rlx\hbox{\rm I\kern-.18em I}}
\def\IK{\rlx\hbox{\rm I\kern-.18em K}}
\def\IL{\rlx\hbox{\rm I\kern-.18em L}}
\def\one{\hbox{{1}\kern-.25em\hbox{l}}}
\def\0#1{\relax\ifmmode\mathaccent"7017{#1}%
B        \else\accent23#1\relax\fi}

\def\EPJC#1#2#3{{\sl Eur. Phys. J.} {\bf C#1} (#2) #3}
                \def\JHEP#1#2#3{{\sl JHEP} {\bf#1} (#2) #3}
                \def\PRL#1#2#3{{\sl Phys. Rev. Lett.} {\bf#1} (#2) #3}
                \def\NPB#1#2#3{{\sl Nucl. Phys.} {\bf B#1} (#2) #3}

                \def\PRD#1#2#3{{\sl Phys. Rev.} {\bf D#1} (#2) #3}
                
                \def\PLA#1#2#3{{\sl Phys. Lett.} {\bf #1A} (#2) #3}
                \def\PLB#1#2#3{{\sl Phys. Lett.} {\bf #1B} (#2) #3}
                \def\JMP#1#2#3{{\sl J. Math. Phys.} {\bf #1} (#2) #3}

                \def\AoP#1#2#3{{\sl Annals Phys.} {\bf #1} (#2) #3}
                
                \def\RMP#1#2#3{{\sl Rev. Mod. Phys.} {\bf #1} (#2) #3}

                \def\JPA#1#2#3{{\sl J. Physics: Math. and Gen.} {\bf A#1} (#2) #3}
                
                \def\MPLA#1#2#3{{\sl Mod. Phys. Lett.} {\bf A#1} (#2) #3}

                \def\JPIV#1#2#3{{\sl J. Phys. IV} {\bf #1} (#2) #3}
                
                \def\PD#1#2#3{{\sl Physica} {\bf D#1} (#2) #3}
                \def\IJPAM#1#2#3{{\sl International J. of Pure and Appl. Math.} {\bf #1} (#2) #3}
\def\JPAMT#1#2#3{{\sl J. Phys. A: Math. Theor.} {\bf #1} (#2) #3}
                \def\a{\alpha}
                \def\b{\beta}

                \def\d{\delta}

                \def\g{\gamma}
                
                \def\vp{\varphi}

                \def\/{\frac}

                \def\l{\lambda}
                \def\L{\Lambda}

                \def\pa{\partial}

                \def\ra{\rightarrow}
                
                \def\vp{\varphi}

                \def\th{\theta}

                \def\({\Big(}
                \def\){\Big)}
                \def\[{\Big[}
                \def\]{\Big]}

                \def\rlx{\relax\leavevmode}
                \def\inbar{\vrule height1.5ex width.4pt depth0pt}
                \def\IZ{\rlx\hbox{\sf Z\kern-.4em Z}}
                \def\IR{\rlx\hbox{\rm I\kern-.18em R}}
                \def\IC{\rlx\hbox{\,$\inbar\kern-.3em{\rm C}$}}
                \def\IN{\rlx\hbox{\rm I\kern-.18em N}}
                \def\IO{\rlx\hbox{\,$\inbar\kern-.3em{\rm O}$}}
                \def\IP{\rlx\hbox{\rm I\kern-.18em P}}
                \def\IQ{\rlx\hbox{\,$\inbar\kern-.3em{\rm Q}$}}
                \def\IF{\rlx\hbox{\rm I\kern-.18em F}}
                \def\IG{\rlx\hbox{\,$\inbar\kern-.3em{\rm G}$}}
                \def\IH{\rlx\hbox{\rm I\kern-.18em H}}
                \def\II{\rlx\hbox{\rm I\kern-.18em I}}
                \def\IK{\rlx\hbox{\rm I\kern-.18em K}}
                \def\IL{\rlx\hbox{\rm I\kern-.18em L}}
                \def\one{\hbox{{1}\kern-.25em\hbox{l}}}
                \def\0#1{\relax\ifmmode\mathaccent"7017{#1}%
                B        \else\accent23#1\relax\fi}
                
\begingroup\makeatletter\ifx\SetFigFont\undefined
\def\x#1#2#3#4#5#6#7\relax{\def\x{#1#2#3#4#5#6}}%
\expandafter\x\fmtname xxxxxx\relax \def\y{splain}%
\ifx\x\y   
\gdef\SetFigFont#1#2#3{%
  \ifnum #1<17\tiny\else \ifnum #1<20\small\else
  \ifnum #1<24\normalsize\else \ifnum #1<29\large\else
  \ifnum #1<34\Large\else \ifnum #1<41\LARGE\else
     \huge\fi\fi\fi\fi\fi\fi
  \csname #3\endcsname}%
\else \gdef\SetFigFont#1#2#3{\begingroup
  \count@#1\relax \ifnum 25<\count@\count@25\fi
  \def\x{\endgroup\@setsize\SetFigFont{#2pt}}%
  \expandafter\x
    \csname \romannumeral\the\count@ pt\expandafter\endcsname
    \csname @\romannumeral\the\count@ pt\endcsname
  \csname #3\endcsname}%
\fi\endgroup

\begin{document}
\pagestyle{plain}
\renewcommand{\thefootnote}{\arabic{footnote}}

\begin{titlepage}

                \begin{center}

                  {\large\bf Noncommutative (generalized) sine-Gordon/massive Thirring 
correspondence, integrability and solitons}

                \end{center}

\vspace{.5 cm}

                \begin{center}

               H. Blas $^{a}$  and  H. L. Carrion $^{b}$ \\

                \vspace{.6 cm}

a) Instituto de F\'{\i}sica,
   Universidade Federal de Mato Grosso\\
   Av. Fernando Correa, s/n, Coxip\'o \\
   78060-900, Cuiab\'a - MT - Brazil\\
b)  Escola de Ci\^encias e Tecnologia / UFRN\\
Campus Universit\'ario Lagoa Nova, CEP 59078-970 - Natal - RN - Brazil.
\end{center}

                \begin{abstract}

                \vspace{.4 cm}
Some properties of the correspondence between the non-commutative versions of the
  (generalized) sine-Gordon (NCGSG$_{1,2}$) and the massive Thirring
   (NCGMT$_{1,2}$) models are studied. Our method relies on the master Lagrangian 
approach to deal with dual theories. The master Lagrangians turn out to be the NC 
versions of the so-called affine Toda model coupled to matter fields (NCATM$_{1,2}$), in 
which the Toda field $g$ belongs to certain subgroups of $ GL(3)$, and the matter fields 
lie in  the higher grading  directions of an affine Lie algebra. Depending on the form of
 $g$ one arrives at two different NC versions of the NCGSG$_{1,2}$/NCGMT$_{1,2}$ correspondence. 
In the NCGSG$_{1,2}$ sectors, through consistent reduction procedures, we find NC versions 
of some well-known models, such as the NC sine-Gordon (NCSG$_{1,2}$) 
(Lechtenfeld et al. and Grisaru-Penati proposals, respectively), NC (bosonized) 
Bukhvostov-Lipatov (NCbBL$_{1,2}$) and NC double sine-Gordon (NCDSG$_{1,2}$) models. The 
NCGMT$_{1,2}$ models correspond to  Moyal product extension of the generalized massive 
Thirring model. The NCGMT$_{1,2}$ models posses constrained versions with relevant Lax 
pair formulations, and other sub-models such as the NC massive Thirring (NCMT$_{1,2}$), 
the NC Bukhvostov-Lipatov (NCBL$_{1,2}$) and constrained versions of the last models 
with Lax pair formulations. We have established that, except for the well known 
NCMT$_{1,2}$ zero-curvature formulations, generalizations 
($n_{F} \ge 2$, $n_F=$number of flavors) of the massive Thirring model allow zero-curvature
 formulations only for constrained versions of the models and for each one of the various 
constrained sub-models defined for less than $n_F$ flavors, in the both NCGMT$_{1,2}$ and 
ordinary space-time descriptions (GMT), respectively. The non-commutative solitons and 
kinks of the $ GL(3)$ NCGSG$_{1,2}$ models are investigated.
\end{abstract}





                \vspace{2 cm}

                \end{titlepage}

\section{Introduction}
\label{intro}

Field theories in non-commutative (NC) space-times are receiving
considerable attention in recent years in connection to the
low-energy dynamics of D-branes in the presence of background
B-field (see e.g. \cite{seiberg}). In particular, the NC
versions of integrable systems (in two dimensions) are being
considered  \cite{hamanaka}. On the other hand, conformal theories on the usual two-dimensional space-time play an important role in various aspects of modern physics, from string theory to applications in condensed matter. So,  one might ask about the role played by
QFT�s in (1 + 1)-dimensional non-commutative space-time. Indeed there is reason to believe that similar applications would emerge and they deserve further investigations, since it is possible to define notions of conformal invariance, Kac-Moody and
Virasoro symmetries in this context \cite{vitale}. Furthermore, there is some optimism regarding the following analogy with the usual known relationship: it is believed that the integrable models, defined on two-dimensional NC {\sl Euclidean} space, would  be the NC versions of statistical models in the critical points and in the off-critical integrable directions.

The sine-Gordon type and other related integrable systems have appeared frequently in diverse areas of physics,
from condensed matter to  string theory, in connection to such properties as soliton solutions, integrability and duality. So, the study of their properties and the search for their solutions have greatly attracted the
 interest of the scientific community. In condensed matter, we can mention for example the work \cite{daniel} on the nonlinear
dynamics of the inhomogeneous DNA double helices chain. In topics of string theory we can mention the
recent works on the magnon-type solutions on the $R\times S^n$ ($n=2,3 $) background geometry \cite{hofman, okamura}.

Some non-commutative versions of the sine-Gordon model (NCSG) have
been proposed in the literature
\cite{grisaru1, grisaru2, jhep2, jhep3, cabrera, lechtenfeld}. The relevant equations of
motion have the general property of reproducing the ordinary
sine-Gordon equation when the non-commutativity parameter is
removed. The Grisaru-Penati version \cite{grisaru1, grisaru2}
introduces a constraint which is non-trivial only in the
non-commutative case. The constraint is required by integrability
but it is satisfied by the one-soliton solutions. However, at the
quantum level this model gives rise to particle production as was
discovered by evaluating tree-level scattering amplitudes
\cite{grisaru2}. On the other hand, introducing an auxiliary
field, Lechtenfeld et al. \cite{lechtenfeld} proposed a novel NCSG
model which seems to possess a factorisable and causal S-matrix.

Recently, in ordinary commutative space the so-called $sl(2)$
affine Toda model coupled to matter (Dirac) fields (ATM) has been
shown to be a Master Lagrangian (ML) from which one can derive the
sine-Gordon and massive Thirring models, describing the
strong/weak phases of the model, respectively
\cite{nucl}-\cite{nucl1}. Besides, the ML approach was
successfully applied in the non-commutative case to uncover
related problems in $(2+1)$ dimensions regarding the duality
equivalence between the Maxwell-Chern-Simons theory (MCS) and the
Self-Dual (SD) model \cite{ghosh}.

In this paper we  extend some properties of the so-called  $sl(3)$ generalized affine Toda model coupled to matter fields (GATM) \cite{jhep1} to the NC case. We define the NCATM model by replacing the products of
fields by the $\star-$products on the level of its effective
action. The effective action associated to this model in ordinary space gives rise to equations of motion
which can be derived from a zero-curvature equation plus some constraints. In fact, the ATM model is a constrained sub-model of an off-critical model related to the so-called conformal affine Toda model coupled to matter fields (CATM) which possesses a Lax pair formulation \cite{matter}.  So, we expect the NCATM model defined in this way does not belong to those class of NC field theories associated to a Lax pair formulation \cite{cabrera}. The NC $GL(2)$ case has been considered in \cite{jhep2}, there the master Lagrangians turn out to be the NC versions of
the ATM model associated to the group
$GL(2)$, in which the Toda field belongs to certain representations of either $U(1) \times U(1)$ or the complexified $U(1)_C$, such that they correspond to the Lechtenfeld et al. (NCSG$_1$) or Grisaru-Penati (NCSG$_{2}$) proposals for the NC versions of the sine-Gordon model, respectively. Besides, the relevant NC massive Thirring (NCMT$_{1,2}$) sectors are written for two (four) types of Dirac fields corresponding to the
Moyal product extension of one (two) copy(ies) of the ordinary massive Thirring model. The NCSG$_{1,2}$ models share the same one-soliton (real Toda field sector of model 2) exact
solutions with their commutative counterparts, which are found without expansion in the NC parameter $\theta$ for the corresponding Toda field. Here the $GL(3)$ extension presents the above known feature regarding the appearance of two versions of the NC (generalized) sine-Gordon model (NCGSG$_{1,2}$) and the corresponding NC  (generalized) massive Thirring models (NCGMT$_{1,2}$), and some new phenomena such as the appearance of the associated sub-models: three copies for each version of the NC sine-Gordon (NCSG$_{1,2}$) models, (bosonized) Bukhvostov-Lipatov models (NCbBL$_{1,2}$), double sine-Gordon models (NCDSG$_{1,2}$),  and three copies for each version of the NC massive Thirring models(NCMT$_{1,2}$), Bukhvostov-Lipatov models (NCBL$_{1,2}$) and the constrained NCBL$_{1,2}$ models, respectively. In addition, we have the known NC soliton solutions in the NCGSG$_{1,2}$ sectors and the appearance of a NC kink type solutions for the NCDSG$_{1,2}$ sub-models. Even though we have discussed the integrability properties of the NCGSG$_{1,2}$ models only for certain integrable directions in field space, i.e. in the NCSG$_{1,2}$ sub-models, the NC generalized massive Thirring (NCGMT$_{1,2}$) sectors present intriguing properties regarding integrability: the NCGMT$_{1,2}$ models encompass a Lax pair formulation only for a sub-model with certain eqs. of motion provided that some constraints are satisfied. Moreover, we established the integrability of certain constrained versions of the NCBL$_{1,2}$ models by providing a corresponding recipe to construct a Lax pair for each of them. The extension of the above features for the $GL(n)$ NCATM$_{1,2}$ models are straightforward.

The study of these models become interesting since the $su(n)$ ATM
theories constitute excellent laboratories to test ideas about
confinement \cite{nucl1, tension}, the role of solitons in quantum
field theories \cite{nucl}, duality transformations interchanging
solitons and particles \cite{nucl, jmp}, as well as the  reduction
processes of the (two-loop) Wess-Zumino-Novikov-Witten (WZNW)
theory from which the ATM models are derivable \cite{matter,
jhep1}. Moreover,  the ATM type systems may also describe some low
dimensional condensed matter phenomena, such as self-trapping of
electrons into solitons, see e.g. \cite{brazovskii}, tunneling in
the integer quantum Hall effect
 \cite{barci}, and, in particular, polyacetylene molecule systems in connection with
 fermion number fractionization \cite{jackiw}. It has been shown
 that the $su(2)$ ATM model describes the low-energy spectrum of QCD$_{2}$ ({\sl one flavor}
 and $N$ colors in the fundamental and $N=2$ in the adjoint representations,
 respectively)\cite{tension}. The $sl(3)$ ATM model and its related dual sub-models GSG/GMT have been used to provide a
 bag model like confinement mechanism for ``quarks"  and it has been shown that the ATM spectrum comprises of solitons as baryons and qualitons as
 constituent quarks in two-dimensional
QCD \cite{jhep4}. Moreover, the $sl(3)$ GSG model has been found to describe the low energy effective action
 of QCD$_{2}$ with unequal 'quark' masses, three flavors and $N$ colors. This model has recently been used to describe the normal and exotic baryon spectrum of QCD$_{2}$ \cite{jhep5}.

The paper is organized as follows. In the next section we present
 the NC extensions of the ATM model relevant to our discussions. It deals with the choice of the group representation  for the Toda field $g$. We introduce two
 types of master Lagrangians (NCATM$_{1,2}$), the {\sl first} one defined for $g \in [U(1)]^{3}$ with the same content of matter fields as the ordinary ATM;
 the {\sl second} one defined for two copies of the NCATM$_{1}$
 such that in this case $g,\, \bar{g} \in {\cal H} \subset SL(3,\IC)$. In section \ref{ncsg12} the non-commutative versions (NCGSG$_{1,2}$)  of the generalized sine-Gordon model (GSG) are derived from
 the relevant master Lagrangians through reduction procedures resembling the one performed in
 the ordinary GATM $\rightarrow$ GSG reduction. In section \ref{parametrizations} we present the two NC extensions (NCGSG$_{1,2}$) of the GSG model, as well as their associated sub-models such as the NCSG$_{1,2}$, NCbBL$_{1,2}$ and NCDSG$_{1,2}$ models.
In section \ref{decoupled} we 'decouple' on shell the
theories NCGSG$_{1,2}$ and
 NCGMT$_{1,2}$, respectively. We discuss the conditions which must satisfy the constraints  in order to have a complete decoupling, in particular for the soliton solutions. In section \ref{ncmt12} we consider the NCGMT$_{1,2}$ models, as well as their global symmetries, associated currents and
 integrability properties of the constrained sub-models. In these developments the double-gauging  of a U(1) symmetry in the
 star-localized Noether procedure to get the currents deserve a careful treatment. We discuss their associated sub-models such as the integrable NCMT$_{1,2}$, the non-integrable NCBL$_{1,2}$, and the (constrained) NCBL$_{1,2}$ models regarded as integrable sub-models.
In section \ref{ncsol} we present the soliton and kink type
 solutions as a sub-set of solutions satisfying the both GSG and NCGSG$_{1,2}$ models  simultaneously. Some discussions and possible directions of research  to pursue in the future  are presented in section \ref{concl}. The Appendix \ref{atm}
  provides the usual GSG model as a reduced $sl(3)$ affine Toda model couple to matter. Some results of the zero-curvature formulation of the CATM model are provided in Appendix  \ref{atmapp}, and the Lagrangian formulation of the ordinary ATM model
 is summarized in Appendix \ref{localatm}.

 \section{The NC affine Toda models coupled to matter fields (NCATM$_{1, 2}$)}
 \label{ncatm}

In this section we present the NC versions of the so-called affine Toda model coupled to
 matter fields (NCATM$_{1, 2}$). The case of $GL(2)$ NCATM model has been studied at the
classical level in \cite{jhep2} and the related NC
sine-Gordon/massive Thirring correspondence has been considered at
the quantum level in \cite{jhep3}. Even though we present detailed computations for the $GL(3)$ case it can follow directly for any $GL(n)$. Two different NC extensions of the ATM model
 (\ref{latm})
 are possible as long as each of them reproduce its ordinary equations of
motion in the commutative limit. The
commutative Toda field $g$ in (\ref{equan1}) belongs to the
complexified abelian subgroup of $SL(3, \IC)$.
 The symmetry group $SL(3)$ of the ordinary ATM model (see
 Appendix \ref{atmapp}) when considered in the NC case is not closed under the Moyal product $\star$; then, the
 NC extension requires the $GL(3)$ group. In the next steps we define
  two versions of the non-commutative $GL(3)$ affine Toda model
 coupled to matter fields (NCATM$_{1,2}$). Let us define the {\sl first} NC extension (NCATM$_{1}$) as
 \begin{eqnarray}
\nonumber
 S_{NCATM_{1}}& \equiv & S[g, W^{\pm},F^{\pm}]  \\\nonumber
 &=& I_{WZW}[g] + \int d^2x \sum_{m=1}^{2}\{\frac{1}{2}<
\partial_{-} W^{-}_{3-m}\star [E_{3}\,,\, W^{-}_{m} ]> - \nonumber\\
&&\frac{1}{2}
< [E_{-3}\,,\,W^{+}_{m}] \star \partial_{+} W^{+}_{3-m} > + <
F^{-}_{m} \star
\partial_{+} W^{+}_{m} > +\nonumber\\&&
<\partial_{-} W^{-}_{m}
                 \star F^{+}_{m}> + <F^{-}_{m}\star g \star F^{+}_{m} \star g^{-1}
                 >\},
               \label{ncatm1} \end{eqnarray}
where $F\star G = F\,
\mbox{exp}\(\frac{\theta}{2}(\overleftarrow{\pa_{+}
}\overrightarrow{\pa_{-}}-\overleftarrow{\pa_{-}}
\overrightarrow{\pa_{+}})\) G $ and  $g \in [U(1)]^3$. In fact, we
have written the NC version of the ATM model presented in the eq. (\ref{latm})
of Appendix \ref{localatm}. The fields $W_{m}^{\pm}, F_{m}^{\pm}$, as well as the generators $E_{\pm 3}$
of the model are defined in eqs. (\ref{fw})-(\ref{ww4}).
$I_{WZW}[g]$ is a NC generalization of the WZNW action for $g$ \br
I_{WZW}[g] = \int d^2x \left[ \pa_{+} g \star \pa_{-} g^{-1} +
\int_0^1 dy \hat{g}^{-1} \star \pa_{y} \hat{g} \star
\[\hat{g}^{-1} \star \pa_{+}  \hat{g}, \hat{g}^{-1} \star \pa_{-}
\hat{g}
\]_{\star}\right], \er
where the homotopy path $\hat{g}(y)$ such that
$\hat{g}(0)=\bf{1}$, $\hat{g}(1) = g$ ($[y,x_{+}]=[y,x_{-}]=0$)
has been defined. The WZW term in this case gives a non-vanishing
contribution due to the non-commutativity. This is in contrast with
the action in ordinary space, i.e. the WZW term in
(\ref{latm})-(\ref{wzw}) vanishes for $g$
belonging to an abelian subgroup of $ SL(3, \IC)$. From (\ref{ncatm1})
one can derive the set of equations of motion for the
corresponding fields \br
  \label{eqmg1}
 \pa_{-}(g^{-1}\star \pa_{+} g)& =& \sum_{m=1}^{2}\[F^{-}_{m}\,,\, g\star F^{+}_{m}\star g^{-1}\]_{\star}
\\
\label{eqmf1}
\partial_{+} F^{-}_{m} &=& [E_{-3}, \partial_{+} W^{+}_{3-m}
],\,\,\,\,\,\,\,\,\,
\partial_{-} F^{+}_{m} = -[E_{3}, \partial_{-} W^{-}_{3-m} ],
 \\
 \partial_{+} W^{+}_{m} &=& - g \star F^{+}_{m} \star g^{-1},\,\,\,\,\,\,\,\,\,\,
  \partial_{-} W^{-}_{m} = - g^{-1} \star F^{-}_{m} \star g \label{eqmw1}.
                \end{eqnarray}

Notice that these set of eqs. closely resemble their commutative counterparts (\ref{atm111})-(\ref{eqmw111}) of Appendix \ref{localatm}. Substituting the derivatives of $W^{\pm}$'s given in the eqs.
(\ref{eqmw1}) into the eqs. (\ref{eqmf1}) one can get the
equivalent set of equations \br \label{eqmf11}
\partial_{+} F^{-}_{m}& =& -[E_{-3}\,,\,  g \star F^{+}_{3-m} \star g^{-1} ],
\,\,\,\,\,\,\,\,\,\,\,
\partial_{-} F^{+}_{m} = [E_{3}\,,\,  g^{-1} \star F^{-}_{3-m} \star g ].
\er

Notice that in the action (\ref{ncatm1}) one can use
simultaneously the cyclic properties of the group trace and the
$\star$ product. Then, the action (\ref{ncatm1}) and the equations
of motion (\ref{eqmg1})-(\ref{eqmw1}) have the left-right local
symmetries given by \br \label{sym1}
g & \rightarrow & h_{L}(x_{-}) \star g(x_{+}, x_{-}) \star  h_{R}(x_{+}),\\
\label{sym2} F^{+}_{m} &\rightarrow& h^{-1}_{R}(x_{+})\star
F^{+}_{m}(x_{+}, x_{-}) \star  h_{R}(x_{+}),\,\,\,\, W^{-}_{m}
\rightarrow h^{-1}_{R}(x_{+})\star W^{-}_{m}(x_{+}, x_{-}) \star
h_{R}(x_{+}),\,\,\,\,\,\,
\\
\label{sym3} F^{-}_{m} &\rightarrow& h_{L}(x_{-})\star
F^{-}_{m}(x_{+}, x_{-})\star h^{-1}_{L}(x_{-}),\,\,\,\,W^{+}_{m}
\rightarrow h_{L}(x_{-})\star W^{+}_{m}(x_{+}, x_{-})\star
h^{-1}_{L}(x_{-}).\er

The system of eqs. (\ref{eqmg1})-(\ref{eqmw1}) is invariant under the
 above symmetries if the following conditions are supplied
\br \label{condi} h_{R}(x_{+})\,\star E_{3}\,
h^{-1}_{R}(x_{+})\,=\, E_{3},\,\,\,\,\, h^{-1}_{L}(x_{-})\,\star
E_{-3}\, h_{L}(x_{-})\,=\, E_{-3}, \er where $h_{L/R}(x_{\mp})\,
\in\, {\cal H}_{0}^{L/R}$, \, ${\cal H}_{0}^{L/R}$ being Abelian
sub-groups of $GL(3)$. These
symmetry transformations written in matrix form \cite{jhep1} are extensions of the ordinary ones to the NC case in a straightforward manner. Notice that in the ordinary space-time, in terms of the field components, the above transformations are given in the Appendix \ref{atm} [see eqs. (\ref{leri1}) and (\ref{leri3})-(\ref{leri4})]; obviously, the form of the expressions given in these eqs. will change in the NC case.

Next, we define the {\sl second} version of the $GL(3)$ NC affine
Toda model coupled to matter NCATM$_{2}$ as \br \label{ncatm2}
S_{NCATM_{2}} \equiv S[g, W^{\pm},F^{\pm}] + S[\bar{g}, {\cal
W}^{\pm},{\cal F}^{\pm}], \er where the independent fields $g$\,
and\, $\bar{g}$, related to the set of matter fields $\{W^{\pm},
F^{\pm}\}$ and $\{ {\cal W}^{\pm},{\cal F}^{\pm}\}$, respectively,
belong to a complexified subgroup \,${\cal H}$\, of $GL(3)$ to be
specified in the subsection \ref{another}. As above the action $S[.\, , \, . \, , \, .]$ is defined as the Moyal extension of (\ref{latm}). The motivation to introduce a copy of the action functional with the set of fields $\bar{g}, {\cal W}^{\pm},{\cal F}^{\pm}$ will be clarified below. Let us mention, in the mean time, that the second version of the NCATM$_{2}$ model has also been considered in \cite{jhep2} for the $SL(2)$ case.  

The equations of motion for the NCATM$_{2}$ model (\ref{ncatm2})
comprise the eqs. (\ref{eqmg1})-(\ref{eqmw1}) written for $g \in
{\cal H} \subset GL(3)$  and
a set of analogous equations for the remaining fields $\bar{g}$,
${\cal F}^{\pm}$ and ${\cal W}^{\pm}$. Moreover, in addition to
the symmetry transformations (\ref{sym1})-(\ref{sym3}) one must
consider similar expressions for $\bar{g}$, ${\cal F}^{\pm}$ and
${\cal W}^{\pm}$.

\section{NC versions of the generalized sine-Gordon model (NCGSG$_{1,2}$)}
\label{ncsg12}

In order to derive the NC versions of the generalized sine-Gordon
model (NCGSG$_{1, 2}$) we follow the master Lagrangian approach
\cite{deser, jhep1}, starting from the NCATM$_{1, 2}$ models
(\ref{ncatm1}) and (\ref{ncatm2}), respectively, as performed in
the $GL(2)$ case \cite{jhep2}. So, let us consider first the
equations of motion (\ref{eqmg1})-(\ref{eqmw1}).
 We proceed by integrating the eqs. (\ref{eqmf1})
\begin{eqnarray}
 \label{solf1}
 F^{-} = [E_{-3},  {W}^{+}_{3-m} ] + f^{-}_{m}(x_{-}),\,\,\,\,
 F^{+} =- [E_{3},  {W}^{-}_{3-m} ] - f^{+}_{m}(x_{+}).
        \end{eqnarray}
with the $f^{\pm}(x_{\pm})$'s being analytic functions. Next, we
replace the $F^{\pm}$ of eqs. (\ref{solf1}) and the $\pa_{\pm}
W^{\pm}$ of (\ref{eqmw1}), written in terms of $W^{\pm}$, into the
action (\ref{ncatm1}) to get
                \begin{eqnarray}
    \nonumber            S'[g, W^{\pm}, f^{\pm}] &=& I_{WZW}[g] +
    \int d^2x \sum_{m=1}^{2}\{ \frac{1}{2}< [E_{-3},W^{+}_{3-m}] \star g
                \star f^{+}_{m} \star g^{-1}> + \\
&& \frac{1}{2}< g^{-1}\star f^{-}_{m}
                \star g \star [E_{3}, W^{-}_{3-m}] > +
                <g^{-1}\star f^{-}_{m} \star g \star f^{+}_{m}> \}.\label{ncsg1aux}
                \end{eqnarray}

As the next step, one writes the equations of motion for the
$f^{\pm}(x_{\pm})$'s and solves for  them; afterwards, substitutes
those expressions into the intermediate action (\ref{ncsg1aux})
getting
 \begin{eqnarray}
                S''[g, W^{\pm}] =  I_{WZW} [g] -
                 \frac{1}{4}\int d^2x \sum_{m=1}^{2} <[E_{-3},W^{+}_{3-m}] \star g \star
                [E_{3},W^{-}_{3-m}] \star g^{-1}>
\label{ncsgeff}.                \end{eqnarray}

Notice that (\ref{ncsgeff}) has inherited from the NCATM action
the local symmetries (\ref{sym1})-(\ref{sym3}). Therefore, one
considers the gauge fixing
                \begin{eqnarray}
\label{gf1}
                2i \Lambda^{-}_{m} = [E_{-3},W^{+}_{3-m}], \;\;\;
                2i \Lambda^{+}_{m}  =[E_{3},W^{-}_{3-m}],
                \end{eqnarray}
where $\Lambda^{\pm} \in \hat{{\cal G}}_{\pm 1}$ are some constant
generators in the subspaces of grade $\pm 1$ in
(\ref{grades2})-(\ref{grades3}).

Then for this gauge fixing the effective action (\ref{ncsgeff}) becomes
                \begin{eqnarray}\nonumber
                S_{NCGSG_{1}}[g] &\equiv & S[g] \\
                &= & I_{WZW} [g] +\int d^2x \sum_{m=1}^{2}[ < \Lambda^{-}_{m}  \star g \star
                \Lambda^{+}_{m}
                \star g^{-1}>].  \label{ncsg1}
                \end{eqnarray}

Thus, we get the equation of motion for the field $g$ as \br
\label{eqncsg1}
 \pa_{-}(g^{-1}\star \pa_{+} g)& =& \sum_{m=1}^{2}\[\L^{-}_{m}\,,\, g \star \L^{+}_{m} g^{-1}\]
\er

The action (\ref{ncsg1}) for  $g \in [U(1)]^3$ will define the first 
version of the non-commutative generalized sine-Gordon model
(NCGSG$_{1}$). The second version requires a copy of the above action for the field $\bar{g}$
\begin{eqnarray}\label{ncgsg22}
                S_{NCGSG_{2}}[g] &\equiv & S[g] + S[\bar{g}], \er
where $g \in {\cal H} \subset SL(3)$ (${\cal H}$ will be specified below).

Thus, the actions (\ref{ncsg1}) and (\ref{ncgsg22}) are the multi-field extensions of the NC sine-Gordon models  proposed earlier by Lechtenfeld et al. and Grisaru-Penati, respectively. As we will see below, these models contain as sub-models the relevant versions of the NCSG$_{1,2}$ model (in fact, each version contains three NCSG$_{1,2}$ sub-models) proposed in the literature, i.e. the Lechtenfeld et al. and Grisaru-Penati proposals for the NC extension of the sine-Gordon model, respectively.  Moreover, the NCGSG$_{1,2}$ models give rise to new phenomena with interesting properties, such as the appearance of two versions of the NC Bukhvostov-Lipatov model and the NC double sine-Gordon model, respectively, as well as their NC soliton and kink type solutions. 

We present below the two NCGSG$_{1,2}$ versions related to $GL(3)$, each one involving multi-field scalar fields. 

\section{The Toda field $g$ parametrizations}
\label{parametrizations}

In this section we present the two possible parametrizations of the field $g$, thus obtaining the two NC versions NCGSG$_{1,2}$ of the GSG model, and furthermore we obtain their relevant sub-models associated to them through consistent reductions.

\subsection{First parametrization: $g \in [U(1)]^3 \subset GL(3,
\IC)$}
\label{ff1}

Let us write the field $g $ in the representation \br
\label{u1u1u1} g = \left(\begin{array}{ccc}
e^{i\phi_{1}}_{\star} & 0&  0 \\
0 & e^{i\phi_{2}}_{\star} & 0 \\
0 & 0 & e^{i\phi_{3}}_{\star}  \\
\end{array} \right)\,\equiv\, g_{1}* g_{2} * g_{3},\, \mbox {where}
\er
  \br \label{factor} g_{1} =
\left(\begin{array}{ccc}
e^{i\phi_{1}}_{\star} &  0 \\
0 & 1 & 0 \\
0 & 0 & 1 \\
\end{array} \right),\,\,\,\, g_{2} = \left(\begin{array}{ccc}
1 &  0 & 0 \\
0 & e^{i\phi_{2}}_{\star}& 0\\
0 & 0 & 1\\
\end{array} \right),\,\,  g_{3} =
\left(\begin{array}{ccc}
0 & 0 & 0 \\
0 & 1 & 0 \\
0 & 0 & e^{i\phi_{3}}_{\star} \\
\end{array} \right)
\er with $\phi_{i}$ being real fields ($i=1,2,3$). As we will see
below, this parametrization constitutes the $GL(3, \IC)$ extension of
the Lechtenfeld et al. proposal of the non-commutative
 version of the  sine-Gordon model (NCSG$_{1}$) \cite{lechtenfeld}.

For the $\Lambda_{i}$'s taken as
\begin{eqnarray}
\Lambda^{+}_{1} &=&  \Lambda^{1}_{R} E^{0}_{\alpha_{1}}+
\Lambda^{2}_{R} E^{0}_{\alpha}+  \tilde{\Lambda}^{3}_{R}
   E^{1}_{-\alpha_{3}}, \nonumber \\
\Lambda^{-}_{1} &=&  \Lambda^{3}_{L} E^{-1}_{\alpha_{3}}+
\tilde{\Lambda}^{1}_{L} E^{0}_{-\alpha_{1}}+ \tilde{\Lambda}^{2}_{L}
   E^{0}_{-\alpha_{2}}, \nonumber \\
   \Lambda^{+}_{2} &=&  \Lambda^{3}_{R} E^{0}_{\alpha_{3}}+
\tilde{\Lambda}^{1}_{R} E^{1}_{-\alpha_{1}}+ \tilde{\Lambda}^{2}_{R}
   E^{1}_{-\alpha_{2}}. \nonumber \\
\Lambda^{-}_{2} &=& \Lambda^{1}_{L} E^{-1}_{\alpha_{1}}+
\Lambda^{2}_{L} E^{-1}_{\alpha_{2}}+\tilde{\Lambda}^{3}_{L}
   E^{0}_{-\alpha_{3}},
   \label{cte3}
   \end{eqnarray}
the action (\ref{ncsg1}) for $g$ given in  (\ref{u1u1u1}), upon
using twice the Polyakov-Wiegmann identity
\br \label{pw}
I_{WZW}(g_{1}*g_{2})= I_{WZW}(g_{1}) + I_{WZW}(g_{2}) +\int dz^2 <
g_{1}^{-1} \star
\partial_{-} g_{1} \star \partial_{+}g_{2} \star g_{2}^{-1}>, \er
can be written as
\begin{eqnarray}\label{nclech12}
S_{NCSG_{1}}[g_{1}, g_{2}, g_{3}] = & &I_{WZW}[g_{1}] +
I_{WZW}[g_{2}] + I_{WZW}[g_{3}] + \nonumber  \\
&& \int d^2x \( [ \Lambda^{3}_{L} \tilde{\Lambda}^{3}_{R}
e_{\star}^{i \phi_{1}} \star e_{\star}^{-i \phi_{3}}+
\tilde{\Lambda}^{3}_{L}
{\Lambda}^{3}_{R} e_{\star}^{i \phi_{3}} \star e_{\star}^{-i \phi_{1}}]+ \nonumber \\
&& [ \Lambda^{1}_{L}  \tilde{\Lambda}^{1}_{R} e_{\star}^{i
\phi_{1}} \star e_{\star}^{-i \phi_{2}}+ \tilde{\Lambda}^{1}_{L}
{\Lambda}^{1}_{R} e_{\star}^{i \phi_{2}} \star e_{\star}^{-i \phi_{1}}]+ \nonumber \\
 && [ \Lambda^{2}_{L}  \tilde{\Lambda}^{2}_{R} e_{\star}^{i \phi_{2}} \star
                   e_{\star}^{-i \phi_{3}}+ \tilde{\Lambda}^{2}_{L}  {\Lambda}^{2}_{R} e_{\star}^{i
                   \phi_{3}}\star
                   e_{\star}^{-i \phi_{2}}] \).
   \end{eqnarray}

Notice that the last term in the Polyakov-Wiegmann identity
(\ref{pw}) vanishes when written for each pair of the fields in
the  parametrizations (\ref{factor}). Then, the relevant  eqs. of
motion become
\br\label{ncsgeq21}
\pa_{-}\(e_{\star}^{-i\phi_{1}}\star \pa_{+} e_{\star}^{i
\phi_{1}}\)&=&  [\Lambda^{1}_{L} \tilde{\Lambda}^{1}_{R}
e_{\star}^{i \phi_{2}} \star e_{\star}^{-i \phi_{1}}-
\tilde{\Lambda}^{1}_{L}{\Lambda}^{1}_{R} e_{\star}^{i \phi_{1}} \star e_{\star}^{-i \phi_{2}}]+ \nonumber \\
&&  [\Lambda^{3}_{L}  \tilde{\Lambda}^{3}_{R} e_{\star}^{i
\phi_{3}} \star e_{\star}^{-i \phi_{1}}-
\tilde{\Lambda}^{3}_{L}{\Lambda}^{3}_{R} e_{\star}^{i \phi_{1}} \star e_{\star}^{-i \phi_{3}}]\\
\pa_{-}\(e_{\star}^{-i\phi_{2}} \star \pa_{+} e_{\star}^{i
\phi_{2}} \)&=&  [\Lambda^{2}_{L}  \tilde{\Lambda}^{2}_{R}
e_{\star}^{i \phi_{3}} \star e_{\star}^{-i \phi_{2}}-
\tilde{\Lambda}^{2}_{L}{\Lambda}^{2}_{R} e_{\star}^{i \phi_{2}} \star e_{\star}^{-i \phi_{3}}]+\nonumber \\
&&  [ \tilde{\Lambda}^{1}_{L}{\Lambda}^{1}_{R} e_{\star}^{i
\phi_{1}} \star e_{\star}^{-i \phi_{2}}-\Lambda^{1}_{L}
\tilde{\Lambda}^{1}_{R} e_{\star}^{i \phi_{2}}
\star e_{\star}^{-i \phi_{1}}].\label{ncsgeq22}\\
\pa_{-}\(e_{\star}^{-i\phi_{3}} \star \pa_{+} e_{\star}^{i
\phi_{3}} \)&=&  [\tilde{\Lambda}^{3}_{L}{\Lambda}^{3}_{R}
e_{\star}^{i \phi_{1}} \star e_{\star}^{-i
\phi_{3}}-\Lambda^{3}_{L} \tilde{\Lambda}^{3}_{R} e_{\star}^{i
\phi_{3}} \star e_{\star}^{-i \phi_{1}}]+ \nonumber \\
&& [ \tilde{\Lambda}^{2}_{L}{\Lambda}^{2}_{R} e_{\star}^{i
\phi_{2}} \star e_{\star}^{-i \phi_{3}}-\Lambda^{2}_{L}
\tilde{\Lambda}^{2}_{R} e_{\star}^{i \phi_{3}} \star e_{\star}^{-i
\phi_{2}}].\label{ncsgeq3} \er

Setting \br \label{cond1} \Lambda^{j}_{L} \tilde{\Lambda}^{j}_{R}
= e^{i \delta_{j}} M_{j}/8, \;\,\,\,j=1,2,3;\er for $M_{j},\,\delta_{j} $ \, some constants, we define the system of eqs. (\ref{ncsgeq21})-(\ref{ncsgeq3}) as the {\sl first version} of the non-commutative generalized $GL(3,
\IC)$ sine-Gordon model (NCGSG$_{1}$). Notice that it is defined for three real scalar fields.

In the commutative limit $\theta
\rightarrow 0$ the above equations can be written as \br
\label{limit1}\pa^2\,\phi_{1}&=&
M_{1}\sin(\phi_{2}-\phi_{1}+\delta_{1})+
M_{3}\sin(\phi_{3}-\phi_{1}+\delta_{3});\\
\pa^2\,\phi_{2}&=&  M_{2} \sin(\phi_{3}-\phi_{2}+\delta_{2})+
M_{1}\sin(\phi_{1}-\phi_{2}-\delta_{1});\label{limit2}\\
\pa^2\,\phi_{3}&=&  M_{2} \sin(\phi_{2}-\phi_{3}-\delta_{2})+
M_{3}\sin(\phi_{1}-\phi_{3}-\delta_{3}).\label{limit3} \er

From the above system of equations one gets a free scalar equation
of motion \br \pa^2\,\Phi&=&0, \,\,\,\,\,\,\Phi \equiv
\phi_{1}+\phi_{2}+\phi_{3}. \label{free0}\er

For the particular solution $\Phi\equiv 0$ of (\ref{free0}) and
making $M_{j} \rightarrow - M_{j},\,\,\,\phi_{1} \rightarrow
-\phi_{1}$, one can write the first two equations
(\ref{limit1})-(\ref{limit2}) as \br
\label{limit111}\pa^2\,\phi_{1}&=&  M_{1}
\sin(\phi_{2}+\phi_{1}+\d_{1})+
M_{3}\sin(2\phi_{1}-\phi_{2}+\d_{3});\\
\pa^2\,\phi_{2}&=&  M_{2} \sin(2\phi_{2}-\phi_{1}-\d_{2})+ M_{1}
\sin(\phi_{1}+\phi_{2} + \d_{1}).\label{limit222} \er

This system of eqs. is precisely the commutative generalized
sine-Gordon model (GSG) \cite{jhep4, jhep5} [the form written in
(\ref{limit111})-(\ref{limit222}) corresponds to eqs. (\ref{sys1})-(\ref{sys2}) of Appendix \ref{atm}].

In the following subsections we will examine certain sub-models obtained through  consistent reductions of the NCGSG$_{1}$  system (\ref{ncsgeq21})-(\ref{ncsgeq3}).

\subsubsection{Non-commutative sine-Gordon model (NCSG$_{1}$): Lechtenfeld et al. proposal}

We show that the model (\ref{ncsgeq21})-(\ref{ncsgeq3})
contains as sub-models the Lechtenfeld et al. proposal for the NCSG$_{1}$ model. So, setting
$M_{2}=M_{3}=0,\,M_{1}=8M,\,\phi_{3}=\d_{j}=0$ and changing
$\phi_{2}\rightarrow -\phi_{2}$ we get the system of equations
\cite{lechtenfeld}\br \label{lechtenfeld1}
\pa_{-}\(e_{\star}^{-i\phi_{1}}\star \pa_{+} e_{\star}^{i
\phi_{1}}\)&=& M [ e_{\star}^{-i \phi_{2}} \star e_{\star}^{-i
\phi_{1}}-
 e_{\star}^{i \phi_{1}} \star e_{\star}^{i \phi_{2}}]\\
 \label{lechtenfeld2}
\pa_{-}\(e_{\star}^{i\phi_{2}} \star \pa_{+} e_{\star}^{-i
\phi_{2}} \)&=& M [  e_{\star}^{i \phi_{1}} \star e_{\star}^{i
\phi_{2}}- e_{\star}^{-i \phi_{2}} \star e_{\star}^{-i \phi_{1}}].
\er In fact, there are additional two possibilities for meaningful
reductions, i.e., 1)\,
 $M_{1}=M_{2}=0,\,M_{3}=8M,\,\phi_{2}=\d_{j}=0$;
$\phi_{3}\rightarrow -\phi_{3}$ and 2)\,
$M_{1}=M_{3}=0,\,M_{2}=8M,\,\phi_{1}=\d_{j}=0$;
$\phi_{3}\rightarrow -\phi_{3}$ respectively, providing in each case a
Lechtenfeld et al. NCSG$_{1}$ model.

\subsubsection{Non-commutative (bosonized) Bukhvostov-Lipatov model (NCbBL$_{1}$). First version}

Another reduction is possible by making $M_{1}=0,\,\,M_2=M_3=-M$, \, and \, $\phi_1 \rightarrow -\phi_1,\,\, $ in the eqs. (\ref{ncsgeq21})-(\ref{ncsgeq3}) followed by the substitution  $\phi_3=\phi_1-\phi_2$. So, one gets the  set of equations
\br\label{ncBL1}
\pa_{-}\(e_{\star}^{\pm i\phi_{a}}\star \pa_{+} e_{\star}^{\mp i
\phi_{a}}\)&=& -\frac{M}{8} \Big[e_{\star}^{i (\phi_1-\phi_{2})} \star e_{\star}^{\pm i \phi_{a}}- e_{\star}^{\mp i \phi_{a}} \star e_{\star}^{-i (\phi_1-\phi_{2})}\Big],\,\,\,\,\,\,\,\,\,\,\,\,\,a=1,2\\
0&=&\pa_{-}\Big[e_{\star}^{i\phi_{1}}\star \pa_{+} e_{\star}^{-i
\phi_{1}}+ e_{\star}^{-i\phi_{2}}\star \pa_{+} e_{\star}^{i
\phi_{2}}+ e_{\star}^{-i(\phi_{1}-\phi_2)}\star \pa_{+} e_{\star}^{i(
\phi_{1}-\phi_2)} \Big],\nonumber \\
\label{ncBL2}
\er
where the upper (lower) signs in (\ref{ncBL1}) correspond to the index $a=1$($2$) for the field $\phi_{a}$. In the commutative limit the eq. (\ref{ncBL2}) becomes trivial, whereas the set of  equations (\ref{ncBL1}) become $\pa^2\phi_1=M \mbox{sin} (2\phi_1-\phi_2)$\,\,\,and\,\, $\pa^2\phi_2=M \mbox{sin} (2\phi_2-\phi_1)$. Defining the new fields $\psi_1=\frac{1}{2}(\phi_1+\phi_2),\,\,\psi_2=\frac{\sqrt{3}}{2}(\phi_1-\phi_2)$ we arrive at the model $\pa^2\psi_1=M \mbox{sin} (\psi_1)  \mbox{cos} (\sqrt{3}\psi_2),\,\,\, \pa^2\psi_2=M \sqrt{3}\mbox{cos} (\psi_1)  \mbox{sen} (\sqrt{3}\psi_2)$. This system of equations is precisely the bosonized form of the so-called Bukhvostov-Lipatov model \cite{ameduri, saleur, lipatov, jmp}. In view of these relationships we define the model (\ref{ncBL1})-(\ref{ncBL2}) as the {\sl first version} of the non-commutative bosonized Bukhvostov-Lipatov model(NCbBL$_{1}$).

\subsubsection{Non-commutative double sine-Gordon model (NCDSG$_{1}$). First version}

The usual double sine-Gordon model (DSG) is defined in terms of
just one scalar field $\phi$ and the potential terms $[cos(\phi)+ cos(2 \phi)]$ in the action. So, we would like to reduce the above
 model in a consistent way in order to get a sub-model defined for just one scalar
 field. Let us take advantage of a particular solution of the free
 field equation (\ref{free0}). So, we consider the reduction
 $ \phi_{1} = -\phi_{3} =
\phi,\;\phi_{2}=0$   and substitute these relations into the
equations (\ref{ncsgeq21})-(\ref{ncsgeq3}). Then we obtain the
next two equations
 \br\label{dsge1}
 \pa_{-}\(e_{\star}^{-i\phi}\star \pa_{+}
e_{\star}^{i \phi}\)&=& M_{1} (   e_{\star}^{-i \phi}-
e_{\star}^{i \phi})+   M_{3} (  e_{\star}^{-i \phi} \star
e_{\star}^{-i \phi} -  e_{\star}^{i
\phi} \star e_{\star}^{i \phi}), \\
\label{dsge2} \pa_{-}\(e_{\star}^{i\phi}\star \pa_{+}
e_{\star}^{-i \phi}\)&=& M_{1} (   e_{\star}^{i \phi}-
e_{\star}^{-i \phi})+   M_{3} ( e_{\star}^{i \phi} \star
e_{\star}^{i \phi} -  e_{\star}^{-i \phi} \star e_{\star}^{-i
\phi})
\end{eqnarray}
plus an equation which reduces to a trivial identity (we have
imposed $M_{1}=M_{2},\, \delta_{i} = 0$).

The above two equations can be written in the equivalent form
\br\label{ncdsg11} \pa_{-}\(e^{i \phi}_{\star} \star \pa_{+} e^{-i
\phi}_{\star} - e_{\star}^{-i\phi}\star \pa_{+} e_{\star}^{i
\phi}\) &=& 4i M_{1}\, \mbox{sin}_{\star} \phi + 4i M_{3}\,
\mbox{sin}_{\star} 2 \phi \\
\pa_{-}\(e^{-i \phi}_{\star} \star \pa_{+} e^{i \phi}_{\star}
+e^{i \phi}_{\star} \star \pa_{+} e^{-i \phi}_{\star}
\)&=&0\label{ncdsg12}.\er

The system (\ref{ncdsg11})-(\ref{ncdsg12}) constitutes the {\sl
first} version of the non-commutative double sine-Gordon model
(NCDSG$_{1}$) defined for just one scalar field.

The first equation (\ref{ncdsg11}) contains the potential terms
which is the natural generalization of the ordinary double
sine-Gordon  potential, whereas the other one (\ref{ncdsg12}) has
the structure of a conservation law and it can be seen as imposing an
extra condition on the system. In the commutative limit, the first
equation reduces to the ordinary double sine-Gordon equation
(DSG), whereas the second one becomes trivial. The equations are
in general complex and possess the $\IZ_{2}$ symmetry of the
ordinary DSG (the invariance under $\phi \rightarrow -\phi$ is
easily seen in (\ref{dsge1})-(\ref{dsge2}).

\subsection{Second parametrization: $ g \in {\cal H} \subset
GL(3, \IC)$} 
\label{another}

Let us consider the parametrization
\br \label{2para} g =
\left(\begin{array}{ccc}
 e^{\vp_{1}}_{\star} \star e^{ \vp_{0}}_{\star}  & 0&  0 \\
0 &  e^{-\vp_{1}+ \vp_{2}}_{\star} \star e^{ \vp_{0}}_{\star} & 0 \\
0 & 0 &  e^{-\vp_{2}}_{\star} \star e^{ \vp_{0}}_{\star}  \\
\end{array} \right)\,\equiv\, g_{1}* g_{2} ,
\er with \br g_{1} = \left(\begin{array}{ccc}
e^{\vp_{1}}_{\star} &  0 & 0 \\
0 & e^{- \vp_{1}+\vp_{2}}_{\star}& 0\\
0 & 0 & e^{-\vp_{2}}_{\star}\\
\end{array} \right),\; g_{2} = e^{ \vp_{0}}_{\star}
\left(\begin{array}{ccc}
1 &  0 \\
0 & 1 & 0 \\
0 & 0 & 1 \\
\end{array} \right),
\er where the fields $\vp_{j},\,j=0,1,2$ are general complex fields. The
additional field $\bar{g}$ is defined by substituting the fields
$\vp_{j}$ above as $\vp_{j}^{\dagger}$. The fields $g$ and
$\bar{g}$ are formally considered to be independent fields.

This parametrization becomes the $GL(3)$ extension of the
Grisaru-Penati proposal for the non-commutative
 version of the sine-Gordon model (NCSG$_{2}$) \cite{grisaru1, grisaru2}.

The following equations of motion can be obtained directly from the first term $S[g]$ of 
the action (\ref{ncgsg22}) for the parametrization (\ref{2para})
\br \pa_{-}\(e_{\star}^{-\vp_{0}}\star e_{\star}^{-\vp_{1}}\star
\pa_{+} \( e_{\star}^{ \vp_{1}}\ \star e_{\star}^{\vp_{0}}\) \)&=&
[\Lambda^{1}_{L} \tilde{\Lambda}^{1}_{R} e_{\star}^{-\vp_{1}+
\vp_{2}} \star e_{\star}^{-\vp_{1}}-
\tilde{\Lambda}^{1}_{L}{\Lambda}^{1}_{R} e_{\star}^{\vp_{1}} \star e_{\star}^{\vp_{1}- \vp_{2}}]+ \nonumber \\
&&  [\Lambda^{3}_{L}  \tilde{\Lambda}^{3}_{R} e_{\star}^{-
\vp_{2}} \star e_{\star}^{- \vp_{1}}-
\tilde{\Lambda}^{3}_{L}{\Lambda}^{3}_{R} e_{\star}^{\vp_{1}} \star e_{\star}^{ \vp_{2}}].\label{ncneweq1} \\
\pa_{-}\(e_{\star}^{-\vp_{0}}\star
e_{\star}^{\vp_{1}-\vp_{2}}\star \pa_{+} \(
e_{\star}^{-\vp_{1}+\vp_{2} }\ \star e_{\star}^{\vp_{0}}\) \)&=&
[\Lambda^{2}_{L} \tilde{\Lambda}^{2}_{R} e_{\star}^{- \vp_{2}}
\star e_{\star}^{\vp_{1}- \vp_{2}}-
\tilde{\Lambda}^{2}_{L}{\Lambda}^{2}_{R} e_{\star}^{-\vp_{1}+\vp_{2}} \star e_{\star}^{ \vp_{2}}]+\nonumber \\
&&  [ \tilde{\Lambda}^{1}_{L}{\Lambda}^{1}_{R} e_{\star}^{
\vp_{1}} \star e_{\star}^{\vp_{1}-\vp_{2}}-\Lambda^{1}_{L}
\tilde{\Lambda}^{1}_{R} e_{\star}^{-\vp_{1}+\vp_{2}}
\star e_{\star}^{-\vp_{1}}]\nonumber \\ \label{ncneweq2}\\
\pa_{-}\(e_{\star}^{-\vp_{0}}\star e_{\star}^{\vp_{2}}\star
\pa_{+} \( e_{\star}^{ -\vp_{2}}\ \star e_{\star}^{\vp_{0}}\)
\)&=& [\tilde{\Lambda}^{3}_{L}{\Lambda}^{3}_{R} e_{\star}^{
\vp_{1}} \star e_{\star}^{ \vp_{2}}-\Lambda^{3}_{L}
\tilde{\Lambda}^{3}_{R} e_{\star}^{-
\vp_{2}} \star e_{\star}^{- \vp_{1}}]+ \nonumber \\
&& [ \tilde{\Lambda}^{2}_{L}{\Lambda}^{2}_{R} e_{\star}^{-\vp_{1}+
\vp_{2}} \star e_{\star}^{ \vp_{2}}-\Lambda^{2}_{L}
\tilde{\Lambda}^{2}_{R} e_{\star}^{-\vp_{2}} \star
e_{\star}^{\vp_{1}- \vp_{2}}]. \nonumber\\
\label{ncneweq3} \er

Introduce the parameters $M_{i}, \d_{i}$ as in (\ref{cond1}). So, we define the system of eqs. (\ref{ncneweq1})-(\ref{ncneweq3}), supplied with the relevant eqs. of motion for the fields $\vp_{j}^{\dagger}$ derived from the second term $S[\bar{g}]$ of the action (\ref{ncgsg22}),  as the {\sl second version} of the non-commutative generalized $GL(3,\IC)$ sine-Gordon model (NCGSG$_{2}$), where the three scalar fields $\vp_{j}$ are in general complex.

Next, let us examine the commutative limit. Redefining $ \vp_{a} \rightarrow  i\, \vp_{a}\,$ (where the new
$\vp_{a}'$s  are real), using definition (\ref{cond1}) and taking
the limit $\theta \rightarrow 0 $ in the above system of equations (\ref{ncneweq1})-(\ref{ncneweq3})
one can get \br
\partial^2 \vp_{1} &=& M_{1} \sin
(2\vp_{1}-\vp_{2}-\delta_{1}) +  M_{3} \sin
(\vp_{1}+\vp_{2}-\delta_{3}) \label{eq1}\\\label{eq2}
\partial^2 \vp_{2} &=& M_{2} \sin
(2\vp_{2}-\vp_{1}-\delta_{2}) +  M_{3} \sin
(\vp_{1}+\vp_{2}-\delta_{3}) \\
\partial^2 \vp_{0} &=&0 \label{free1}.\er

Thus, in (\ref{eq1})-(\ref{eq2}) we recover again the equations of
motion of the commutative generalized sine-Gordon model (GSG)
\cite{jhep4, jhep5}. Notice that the field $\vp_{0}$ decouples
completely from the other fields in this limit, becoming
 simply a free field.

In analogy to the results of the first parametrization it is possible to get some sub-models as consistent reductions of the system (\ref{ncneweq1})-(\ref{ncneweq3}). In the following we discuss the reductions associated to this second parametrization.

\subsubsection{Non-commutative sine-Gordon model (NCSG$_{2}$): Grisaru-Penati proposal}

A reduced single field model follows by setting $M_{2}=M_{3}=\d_{i} =0$, $\vp_{0}=\vp_{2}=0$, $M_{1}=-M$ and $\vp_{1}=i\vp$ ($\vp$, complex field). So, one gets the model
\br\label{grisaru1} \pa_{-}(e^{\mp i\vp} \pa_{+} e^{\pm i\vp})= \pm M (e^{\pm 2i\vp}-e^{\mp 2i\vp}), \er
which is the Grisaru-Penati proposal for the NC extension of the sine-Gordon model (NCSG$_{2}$) \cite{grisaru1, grisaru2}. In fact, in this proposal one must consider  additionally a couple of equations for $\vp^{\dagger}$ obtained from the second piece in the action (\ref{ncgsg22}).  Additional reductions, each one providing a Grisaru-Penati  NCSG$_{2}$ model, are  achieved by setting $M_{1}=M_{3}=\d_{i}=0$, $\vp_{0}=\vp_{1}=0$, $M_{2}=-M$\,$\vp_{2}=i\vp$,\, and $M_{1}=M_{2}=\d_{i}=0$, $\vp_{0}=0$, $M_{3}=-M$\,$\vp_{1}=\vp_{2}=i\vp$, respectively.

\subsubsection{Non-commutative (bosonized) Bukhvostov-Lipatov model (NCbBL$_{2}$). Second version}

A reduction leading to a two field model follows as $M_3=0,\,\,M_1=M_2=-M,\,\,\,\vp_{n}\rightarrow i\vp_n$ ($n=0,1,2$). So, one gets the model
\br
\pa_{-}\(e_{\star}^{ -i\vp_{0}} \star e_{\star}^{\mp i\vp_{a}}\star \pa_{+} ( e_{\star}^{\pm i \vp_{a}} \star e_{\star}^{ i\vp_{0}})\)&=& -\frac{M}{8} \Big[e_{\star}^{-i (\vp_1-\vp_{2})} \star e_{\star}^{\mp i \vp_{a}}- e_{\star}^{\pm i \vp_{a}} \star e_{\star}^{i (\vp_1-\vp_{2})}\Big],\nonumber \\&& \,\,\,\,\,a=1,2  \label{ncBL11}\\
0&=&\pa_{-}\Big[e_{\star}^{ -i\vp_{0}} \star e_{\star}^{-i\vp_{1}}\star \pa_{+} ( e_{\star}^{ i \vp_{1}} \star e_{\star}^{ i\vp_{0}})+e_{\star}^{ -i\vp_{0}} \star e_{\star}^{ i\vp_{2}}\star \nonumber \\
&&\pa_{+} ( e_{\star}^{-i \vp_{2}} \star e_{\star}^{ i\vp_{0}} ) + e_{\star}^{ -i\vp_{0}} \star e_{\star}^{ i(\vp_1-\vp_{2})}\star \nonumber \\
&& \pa_{+} ( e_{\star}^{-i (\vp_1-\vp_{2})} \star e_{\star}^{ i\vp_{0}} )\Big],  \label{ncBL21}
\er
where the upper (lower) signs in (\ref{ncBL11}) correspond to the index $a=1$($2$) of the field $\vp_{a}$. In the commutative limit the eq. (\ref{ncBL21}) reduces to a free scalar field equation of motion $\pa^2 \vp_0=0$, whereas the set of  equations (\ref{ncBL11}) become $\pa^2\vp_1=M \mbox{sin} (2\vp_1-\vp_2)$\,\,\,and\,\, $\pa^2\vp_2=M \mbox{sin} (2\vp_2-\vp_1)$. Defining the new fields $\psi_1=\frac{1}{2}(\vp_1+\vp_2),\,\,\psi_2=\frac{\sqrt{3}}{2}(\vp_1-\vp_2)$ we arrive at the model $\pa^2\psi_1=M \mbox{sin} (\psi_1)  \mbox{cos} (\sqrt{3}\psi_2),\,\,\, \pa^2\psi_2=M \sqrt{3}\mbox{cos} (\psi_1)  \mbox{sen} (\sqrt{3}\psi_2)$. As we have seen before this is just the bosonized form of the so-called Bukhvostov-Lipatov model \cite{ameduri, saleur, lipatov, jmp}. In view of these relationships we define the model (\ref{ncBL11})-(\ref{ncBL21}) as the {\sl second version} of the non-commutative (bosonized) Bukhvostov-Lipatov model(NCbBL$_{2}$).

\subsubsection{Non-commutative double sine-Gordon model (NCDSG$_{2}$). Second version}

In order to reduce the NCGSG$_{2}$ system of equations into another version of the NC double sine-Gordon model one takes
advantage of certain properties of its commutative counterpart. In
fact, the above commutative model (\ref{eq1})-(\ref{free1})
possesses the symmetry $\vp_{1} \leftrightarrow \vp_{2}; \,\, \,
M_{1} \leftrightarrow M_{2}$ in the GSG sector, whereas the
auxiliary $\vp_{0}$ field completely decouples in this limit. So,
in the second parametrization case (\ref{2para}) we can impose the
conditions $\varphi_{1}=\varphi_{2} \equiv  -i \varphi,\,\,
M_{1}=M_{2}, \delta_{i}=0$ into the system of eqs.
(\ref{ncneweq1})-(\ref{ncneweq3}) and obtain the following system
of equations for complex $\vp$ \br \pa_{-}\(e_{\star}^{-\vp_{0}}\star e_{\star}^{i
\vp}\star \pa_{+} \( e_{\star}^{ -i \vp}\ \star
e_{\star}^{\vp_{0}}\) \)&=& 2i M_{1}\, \mbox{sin}_{\star}\, \vp +
2i M_{3}\, \mbox{sin}_{\star}\,
2 \vp \nonumber\\ \label{ncneweq21} \\
\pa_{-}\(e_{\star}^{-\vp_{0}}\star \pa_{+}  e_{\star}^{\vp_{0}}
\)&=&
0 \label{ncneweq22}\\
\pa_{-}\[e_{\star}^{-\vp_{0}}\star e_{\star}^{i \vp}\star \pa_{+}
\( e_{\star}^{ -i \vp}\ \star e_{\star}^{\vp_{0}}\) +
e_{\star}^{-\vp_{0}}\star e_{\star}^{-i\vp}\star \nonumber\\
\pa_{+} \(
e_{\star}^{ i \vp}\ \star e_{\star}^{\vp_{0}}\)
   \]&=& 0. \label{ncneweq23} \er

The system (\ref{ncneweq21})-(\ref{ncneweq23}) constitutes the
{\sl second} version of the non-commutative double sine-Gordon
model (NCDSG$_{2}$) defined for two complex scalar fields.

The first equation (\ref{ncneweq21}) contains the potential terms
generalizing the ordinary double sine-Gordon potential. The second
and third ones (\ref{ncneweq22})-(\ref{ncneweq23}) have the
structure of conservation laws and can be seen as imposing extra
conditions on the system. Let us examine the commutative limit
$\theta \rightarrow 0$ of the NCDSG$_{2}$ system. In this limit it
reduces to the usual DSG model plus a free field $\vp_{0}$
equations of motion  \br \pa_{-} \pa_{+} \vp &=& -2
M_{1}\, \mbox{sin}\, \vp -2 M_{3}\, \mbox{sin}\, 2 \vp \\
\pa_{-}\pa_{+} \vp_{0}&=& 0. \er Notice that in this limit the
field $\vp_{0}$ decouples completely from the DSG field $\vp$.

Some comments are in order here.

1) The NC models obtained above reproduce the usual models in the
commutative limit $\theta \rightarrow 0$. So, the both versions of
the $GL(3,\IC)$ non-commutative generalized sine-Gordon model
(NCGSG$_{1,\,2}$) reproduce the ordinary $GL(3)$ GSG model in this
limit. The both versions of the non-commutative double sine-Gordon
model NCDSG$_{1,\,2}$ reproduce the usual DSG model in the
ordinary space. Likewise, the both versions of the non-commutative
bosonized Bukhvostov-Lipatov model NCbBL$_{1,\,2}$ lead to the
usual BL model. Notice that the GSG model in ordinary space-time
also contains as sub-models the variety of theories we have
uncovered above, i.e. the usual SG model, Bukhvostov-Lipatov
model, and  the double sine-Gordon model \cite{jmp, jhep4}.

2) Regarding the integrability of the NCGSG$_{1,2}$  models they
are hardly expected to possess  this property since they contain
as sub-models the relevant NCDSG$_{1,2}$ and NCBL$_{1,2}$ theories.
The NCDSG$_{1,2}$ models are not expected to posses this property
since their commutative counterpart is not integrable. The same
behavior may be expected for the NCBL$_{1,2}$ models since their
commutative counterpart is not classically integrable (see
\cite{jmp} and refs. therein), except for some restricted region in
parameters space. Nevertheless, see more on this point in
subsection \ref{NCGMTsubmodels} when the relevant spinor version
of the (constrained) NCBL$_{1}$ model is discussed in relation to
integrability.

Related to this issue, let us mention that we have not been able
to write in a  zero-curvature form the eq. of motion (\ref{eqncsg1}) of the 
NCGSG$_{1}$ model (\ref{eqncsg1}), it mainly happens due to the
presence of the summation index $m=1,2$ on both entries of the
commutator. Actually, the eq. (\ref{eqncsg1}) differs from the
integrable system of non-abelian affine Toda equations
\cite{ferreira, matter}.

3) The Letchenfeld et al.
(\ref{lechtenfeld1})-(\ref{lechtenfeld2}) and  Grisaru-Penati
(\ref{grisaru1}) NC sine-Gordon models proposed in the literature
appear in the context of the generalized NC sine-Gordon  models as
reduced sub-models of the corresponding NCGSG$_{1}$ and NCGSG$_{2}$
models, respectively. So, they are analogous to the results
obtained in the commutative case in which the $GL(3)$ GSG model
contains three SG sub-models as reduced models, each one associated
to the positive root of the $gl(3)$ Lie algebra\cite{jhep4}. The
group structure of the  $GL(3)$ NCGSG$_{1,2}$ models allowed us to get three NCSG$_{1,2}$ sub-models, respectively, for each version, as in the commutative
case.

4) In the three-field space of the NCGSG$_{1,\,2}$  models it is
remarkable the appearance of three integrable directions as
NCSG$_{1,\,2}$ sub-models, respectively. It suggests that there are
at least three integrable directions in reduced field space of
each one of the NCGSG$_{1, \,2}$ models . Examples of
non-integrable reduced directions are provided by the relevant
NCDSG$_{1,\,2}$ and NCBL$_{1,\,2}$ models. However, the existence of more  
integrable directions is suggested by the presence of certain integrable sub-models in 
the spinor sector  of the NCGMT$_{1,2}$ models, i.e. the scalar duals of the corresponding NC(c)GMT$_{1,2}$ and NC(c)BL$_{1,2}$ spinor models, respectively (see section (\ref{zerocurvature}) and subsection (\ref{NCGMTsubmodels})). 

5) Finally, the role played by the SG model in the context of the generalized SG models is analogous to the one which happens with the correspondence between the $\l \phi^4$ model and the deformed linear $O(N)$-sigma model, as it was first noticed in \cite{guilarte}. It could be interesting to study several properties of the generalized SG models, including their non-commutative counterparts, as for example  by applying and improving the quantization method described in the last reference. Let us mention that the ordinary DSG model has been in the
center of some controversy regarding the computation of its
semi-classical spectrum \cite{mussardo, takacs}, however it seems that the point has recently been clarified in \cite{takacs1}.   

\section{Decoupling of  NCGSG$_{1,2}$ and NCGMT$_{1,2}$ models}
\label{decoupled}

In the commutative case some approaches have been proposed in
order to recover the  GSG and GMT dual models out of  the ordinary
$sl(n)$ ATM model \cite{jhep4, nucl, annals, jhep1,jmp}. Among
them, the one which proceeds by decoupling the set of equations of
motion of the ATM model into the corresponding dual models
\cite{nucl, jhep1}  has turned out to be more suitable in the NC
case \cite{jhep2}. This procedure is adapted to the NC case by
writing a set of mappings between the fields of the model such
that the eqs. (\ref{eqmg1}) and (\ref{eqmf11}) when rewritten
using those mappings decouple the scalar and the matter
fields. So, following the procedures employed in the ordinary
sl(n) case \cite{jhep1} and in the  non-commutative $GL(2)$ ATM
case \cite{jhep2} to the case at hand, let us consider the
mappings \br \label{map1} \sum_{n=1}^{2} \[F_{n}^{-}\,,\,
gF_{n}^{+}g^{-1}\]_{\star}  &= &
\sum_{n=1}^{2}\[\L_{n}^{-}\,,\, g\L_{n}^{+}g^{-1}\]_{\star} ,\\
\label{map2}
\[E_{-3}\,,\, gF_{3-m}^{+}g^{-1}\]_{\star} &= &
\[E_{-3}\,,\,F_{3-m}^{+}\]_{\star} - \frac{k_2}{2} (L^{+}_{m})^{-1} \[\sum_{n}  \hat{J}^{-}_{n}\,,\,
\hat{F}_{m}^{-} \]_{\star} L^{+}_{m},\\
\[E_{3}\,,\, g^{-1}F_{3-m}^{-} g \]_{\star} &= &  \[E_{3}\,,\,F_{3-m}^{-}\]_{\star}
-\frac{k_1}{2} (L^{-}_{m})^{-1} \[\sum_{n}  \hat{J}^{+}_{n}\,,\,
\hat{F}_{m}^{+}
\]_{\star}L^{-}_{m}, \label{map3}
\\
\label{map4}
F^{\pm}_{m}&=& \mp  [E_{\pm 3}\,,\, W_{3-m}^{\mp}]_{\star},\\
\label{sumconst1} \[F_{2}^{\pm}\,,\,
g^{\mp 1}F_{1}^{\mp}g^{\pm 1}\]_{\star}&=&0,\er
where
\br
\label{currpm} \hat{J}^{\mp}_{n} &  \equiv &
\[\hat{F}_{3-n}^{\pm}\,,\,\hat{W}_{3-n}^{\mp}\];\,\,\,\,\,\,k_1,k_2 = \mbox{constant parameters}. \er

The hatted fields have the same algebraic structure as the
corresponding unhatted ones except that they incorporate some
parameters re-scaling the fields, those parameters will give rise
to certain coupling constants between the currents of the model.
Notice that the fields $\hat{J}_{m}^{\pm}$ and the constant
matrices $L^{\pm}_{m}$ carry zero gradation and these will be
defined below. The field $g$ in the relations above, as defined in
section 2, is assumed to belong to either $[U(1)]^3$, as in
subsection \ref{ff1}, or ${\cal H} \subset GL(3, \IC)$ , as in the
second parametrization in subsection \ref{another}.

The relationships (\ref{map1})-(\ref{map4}) when
conveniently substituted into the ATM eqs. of motion (\ref{eqmg1})
and (\ref{eqmf11}) decouple them, respectively, into the NCGSG$_{1}$ eq. (\ref{eqncsg1}) and
certain equations of motion incorporating only matter fields, which in matrix form become
\begin{eqnarray}
\label{eqt11}
 \[ E_{-3}, \partial_{+}{W}^{+}_{3-m} \]_{\star} &= & + [E_{-3},
[E_{3},{W}^{-}_{m}]]_{\star} - \nonumber \\
& &  \frac{k_2}{2} \sum^{2}_{n=1}  (L_{m}^{+})^{-1} [\hat{J}^{-}_{n}, [E_{-3},\hat{W}^{+}_{3-m}]]_{\star}L^{+}_{m}  \\
\label{eqt22} \[ E_{3}, \partial_{-} W^{-}_{3-m} \]_{\star} & = &
- [E_{3},
[E_{-3},{W}^{+}_{m}]]_{\star} - \nonumber \\
& &  \frac{k_1}{2} \sum^{2}_{n=1} (L_{m}^{-})^{-1}
[\hat{J}^{+}_{n}, [E_{3},\hat{W}^{-}_{3-m}]]_{\star} L^{-}_{m}
\end{eqnarray}

We define these set of eqs. as the {\sl first version} of the
non-commutative (generalized) massive Thirring model (NCGMT$_{1}$).

The eqs. (\ref{sumconst1}) are the constraints imposed in ref.
\cite{jhep1} written in a compact form. These constraints, which
are missing in the $GL(2)$ case, have been imposed in the non-trivial
$GL(3)$ extension in order to be able to write a local Lagrangian
for the off-critical and constrained ATM model out of the full set of equations of motion of the so-called conformal affine Toda model coupled to matter (CATM) \cite{jhep1, matter} (see the Appendices). Actually, the above 'decoupling' eqs. maintain the
same form as their commutative analogs presented in eqs.
(6.1)-(6.5) of the ref. \cite{jhep1}. We must clarify that the
above 'decoupling' eqs. (\ref{map1})-(\ref{map3}) do not
completely decouple the scalar fields from the spinor-like fields due to the presence of the constraints (\ref{sumconst1}).
There are some instances of total decoupling, e.g. in the soliton
sector of the commutative limit \cite{jmp, jhep1}. Notice that we
have not used the constraint equations (\ref{sumconst1}) in
order to get the eqs. (\ref{eqt11})-(\ref{eqt22}). In order to be
more specific in the discussions below we provide, in the
following set of equations, the constraint eqs. (\ref{sumconst1})
in terms of the component fields. Let us take the spinors as
defined in (\ref{fw})-(\ref{ww4}) and the scalar field $g$
presented in the first parametrization eq. (\ref{u1u1u1}), so one
has  \br e^{i\phi_{1}}_{\star}\star \psi_{R}^{1}\star
e^{-i\phi_{2}}_{\star} \star \psi_{L}^{2}&=& \psi_{L}^{1}\star
e^{i\phi_{2}}_{\star} {\star} \psi_{R}^{2}\star
e^{-i\phi_{3}}_{\star},\label{conscomp1}\\
e^{i\phi_{2}}_{\star}\star
\psi_{R}^{2}\star e^{-i\phi_{3}}_{\star} \star
\widetilde{\psi}_{L}^{3}&=& - \psi_{L}^{2}\star
e^{i\phi_{3}}_{\star}{\star} \widetilde{\psi}_{R}^{3}\star
e^{-i\phi_{1}}_{\star}, \label{conscomp12}\\ e^{i\phi_{3}}_{\star}\star
\widetilde{\psi}_{R}^{3}\star e^{-i\phi_{1}}_{\star} \star
\psi_{L}^{1}&=& -\widetilde{\psi}_{L}^{3}\star e^{i\phi_{1}}_{\star}
{\star}  \psi_{R}^{1}\star
e^{-i\phi_{2}}_{\star} \label{conscomp2}\er 
and 
\br
e^{-i\phi_{3}}_{\star}\star \widetilde{\psi}_{L}^{2}\star
e^{i\phi_{2}}_{\star} \star \widetilde{\psi}_{R}^{1}&=&
\widetilde{\psi}_{R}^{2}\star e^{-i\phi_{2}}_{\star} {\star}
\widetilde{\psi}_{L}^{1}\star
e^{i\phi_{1}}_{\star},\label{conscomp11}\\
e^{-i\phi_{1}}_{\star}\star \psi_{L}^{3}\star
e^{i\phi_{3}}_{\star} \star \widetilde{\psi}_{R}^{2}&=&
-\psi_{R}^{3}\star e^{-i\phi_{3}}_{\star}{\star}
\widetilde{\psi}_{L}^{2}\star
e^{i\phi_{2}}_{\star}, \label{conscomp1122}\\e^{-i\phi_{2}}_{\star}\star
\widetilde{\psi}_{L}^{1}\star e^{i\phi_{1}}_{\star} \star
\psi_{R}^{3}&=&- \widetilde{\psi}_{R}^{1}\star
e^{-i\phi_{1}}_{\star}{\star}
\psi_{L}^{3}\star
e^{i\phi_{3}}_{\star},
\label{conscomp22}\er 
associated to the
grades $(-1)$ and $(+1)$ of (\ref{sumconst1}), respectively.

Analogously, one can write another set of equations for the second
parametrization (\ref{2para}) of $g$ \br e^{\vp_{1}}_{\star}\star
e^{\vp_{0}}_{\star}\star\psi_{R}^{1}\star
e^{-\vp_{0}}_{\star}\star e^{\vp_{1}-\vp_{2}}_{\star} \star
\psi_{L}^{2}&=& \psi_{L}^{1}\star e^{\vp_{2}-\vp_{1}}_{\star}
\star e^{\vp_{0}}_{\star}\star \psi_{R}^{2}\star
e^{-\vp_{0}}_{\star}\star
e^{\vp_{2}}_{\star},\label{conscomponent1}\\
e^{\vp_{2}-\vp_{1}}_{\star}\star
e^{\vp_{0}}_{\star}\star\psi_{R}^{2}\star
e^{-\vp_{0}}_{\star}\star e^{\vp_{2}}_{\star} \star
\psi_{L}^{3}&=& -\psi_{L}^{2}\star e^{-\vp_{2}}_{\star} \star
e^{\vp_{0}}_{\star}\star \widetilde{\psi}_{R}^{3}\star
e^{-\vp_{0}}_{\star}\star e^{-\vp_{1}}_{\star},
 \label{conscomponent2}\\
 e^{-\vp_{2}}_{\star}\star
e^{\vp_{0}}_{\star}\star\widetilde{\psi}_{R}^{3}\star
e^{-\vp_{0}}_{\star}\star e^{-\vp_{1}}_{\star} \star
\psi_{L}^{1}&=&- \widetilde{\psi}_{L}^{3}\star e^{\vp_{1}}_{\star}
\star e^{\vp_{0}}_{\star}\star \psi_{R}^{1}\star
e^{-\vp_{0}}_{\star}\star
e^{\vp_{1}-\vp_{2}}_{\star},\label{conscomponent3}\er and \br
e^{\vp_{0}}_{\star}\star\widetilde{\psi}_{R}^{2}\star
e^{-\vp_{0}}_{\star}\star e^{\vp_{1}-\vp_{2}}_{\star} \star
\widetilde{\psi}_{L}^{1}\star e^{\vp_{1}}_{\star}  &=&
e^{\vp_{2}}_{\star} \star \widetilde{\psi}_{L}^{2}\star
e^{\vp_{2}-\vp_{1}}_{\star} \star e^{\vp_{0}}_{\star}\star
\widetilde{\psi}_{R}^{1}\star
e^{-\vp_{0}}_{\star},\label{conscomponent11}\\
e^{\vp_{0}}_{\star}\star\psi_{R}^{3}\star
e^{-\vp_{0}}_{\star}\star e^{\vp_{2}}_{\star} \star
\widetilde{\psi}_{L}^{2}\star e^{\vp_{2}-\vp_{1}}_{\star}  &=&-
e^{-\vp_{1}}_{\star} \star \psi_{L}^{3}\star e^{-\vp_{2}}_{\star}
\star e^{\vp_{0}}_{\star}\star \widetilde{\psi}_{R}^{2}\star
e^{-\vp_{0}}_{\star},\label{conscomponent22}\\
e^{\vp_{0}}_{\star}\star\widetilde{\psi}_{R}^{1}\star
e^{-\vp_{0}}_{\star}\star e^{-\vp_{1}}_{\star} \star
\psi_{L}^{3}\star e^{-\vp_{2}}_{\star}  &=&-
e^{\vp_{1}-\vp_{2}}_{\star} \star \widetilde{\psi}_{L}^{1}\star
e^{\vp_{1}}_{\star} \star e^{\vp_{0}}_{\star}\star
\psi_{R}^{3}\star e^{-\vp_{0}}_{\star},\label{conscomponent33}
 \er
associated to the grades $(-1)$ and $(+1)$ of (\ref{sumconst1}),
respectively.

Even though that the full set of the 'decoupling' equations have
not been used in order to write the eqs.
(\ref{eqt11})-(\ref{eqt22}), we expect that a non-commutative
version of the usual (generalized) massive Thirring model
(GMT$_{1}$) \cite{jhep1} defined for the fields $W^{\pm}$ will
emerge from these equations. In fact, we assume this point of view
and study the properties of the system (\ref{eqt11})-(\ref{eqt22})
in its own right. Nevertheless, we will recognize below certain
relationships between the relevant sub-models of the both NCGSG$_{1,2}$
and NCGMT$_{1,2}$ sectors. Remarkably, these relationships will arise for
certain reduced sectors obtained such that the constraints
(\ref{sumconst1}) become trivial, or completely decouple the
spinors from the scalars in the soliton sector, which is equivalent to take the   commutative limit (see below). The model
(NCGMT$_{1}$) (\ref{eqt11})-(\ref{eqt22}) is new in the literature
and it is expected to correspond to the weak coupling sector
of the NCATM$_{1}$ model whose strong coupling sector is
described by the first version of the non-commutative generalized
sine-Gordon model (NCGSG$_{1}$) presented in subsection \ref{ff1}.

In the ordinary space the GMT equations of motion can be
achieved through Hamiltonian reduction procedures, such as the
Faddeev-Jackiw method, as employed in \cite{jhep1} for first order
in time Lagrangian; however, in the NC case, to our knowledge,
 there is no a similar procedure since the
action of the NC GATM model involves higher order in time
derivatives; actually, an infinite number of terms of increasing
order in time derivatives. So, we have used the decoupling method and assumed the forms of the
decoupling equations (\ref{map1})-(\ref{currpm}) to resemble the
ones in the ordinary case \cite{jhep1}, important guiding lines being the
gradation structure and further, the locality of the Lagrangian in
the NCGMT sector which will depend on the nature of the terms
appearing in the eqs. of motion; e.g,  notice the absence of terms
bilinear in the spinors in the right hand side of the eqs.
(\ref{eqt11})-(\ref{eqt22}). In fact, the terms appearing in the
above equations will give rise to usual kinetic and mass terms,
and four-spinor coupling terms in the relevant action. The Lagrangian for the model (\ref{eqt11})-(\ref{eqt22}) and a Lax pair
formulation for a constrained version of it will be discussed below.

In order to recover the dual of the second version NCGSG$_{2}$ one must
 write similar decoupling expressions for the full set of fields $\{g, F^{\pm}, W^{\pm} \}$\, and\,
 $\{\bar{g}, {\cal F}^{\pm}, {\cal W}^{\pm}\}$. Thus,
 following similar steps to the previous construction
  we expect to recover another version of the NC generalized massive Thirring model (NCGMT$_{2}$) defined for the
  fields $\{W^{\pm}, {\cal W}^{\pm}\}$. In the next section we
  propose two versions of the non-commutative (generalized) massive Thirring theories (NCGMT$_{1,2}$) by
  providing the relevant equations of motion and discussing their zero-curvature formulations.

\section{The NC generalized massive Thirring models NCGMT$_{1, 2}$}
\label{ncmt12}

We will consider the fields $\psi^{j}, \widetilde{\psi}^{j}$ as
c-number ones \cite{jhep2} in order to define the NC
generalization of the so-called (c-number) massive Thirring model
(MT) \cite{orfanidis, garbaczewski}. In ordinary space-time these
type of classical c-number multi-field  massive Thirring theories
have long been considered in relation to one-dimensional Dirac
model of extended particles \cite{ranada}. The quantization of the
two-dimensional fermion model with Thirring interaction among N
different massive Fermi field species has recently been performed in the functional integral
approach \cite{belvedere}.

The assumption for the fields to be c-number fields will allow the
zero-curvature formulations of the NCGMT$_{1,2}$ models to be
constructed resembling analogous  algebraic structures present in
the GATM model in the context of the affine Lie algebra $SL(3)$.
This means that the c-number fields $\psi^{j},
\widetilde{\psi}^{j}$ will lie in certain higher grading
directions of the principal gradation of the affine $SL(3)$ Lie
algebra, as it is presented in the eqs. (\ref{ww1})-(\ref{ww4}) of
the Appendix \ref{atmapp}.

In ordinary space the field components of the MT model are considered  to be either anti-commuting Grassmannian fields or some ordinary commuting fields (see \cite{jhep2} and refs. therein). Notice that the
relevant (Grassmannian) GMT model would need a slightly different algebraic formulation from the one followed here for the c-number case.

\subsection{NCGMT$_{1}$}
\label{ncmt1s}

We propose the NCGMT$_{1}$ action related to the eqs. of motion
(\ref{eqt11})-(\ref{eqt22}) for the fields $W^{\pm}_{m}$ as
\begin{eqnarray} \label{ncmt1}
S[W^{\pm}_{m} ] &=&  \int dx^2 \Big[  \sum^{2}_{m=1} \{
\frac{1}{2} <[ E_{-3}, W^{+}_{3-m}] \star \partial_{+} W^{+}_{m}>
- \frac{1}{2}< [ E_{3},
W^{-}_{3-m}] \star \partial_{-} W^{-}_{m}> - \nonumber \\
& &  < [E_{-3},W^{+}_{m}] \star [E_{3},W^{-}_{m}]> \} -\frac{1}{2}
\sum^{2}_{m,n=1} <\hat{J}^{+}_{m} \star \hat{J}^{-}_{n}>\Big].
\end{eqnarray}

In the last action the first two terms inside the summation provide the kinetic terms, the third one the mass terms and the last term the current-current interactions. The current-like
 matrices  $\hat{J}^{\pm}_{m}$ with zero gradation appearing in the eq. (\ref{currpm})
have the same algebraic structure as the matrix-valued currents \cite{jhep1}
\begin{eqnarray}
\label{curr1}
 J^{\pm}_{m} &=& \pm \frac{1}{4}  [[E_{\mp 3}, W^{\pm}_{m}],
W^{\pm}_{3-m}]_{\star},
\end{eqnarray}
except that they are defined in terms
of some hatted variables $\hat{W}_{m}^{\pm}$ which are constructed
 from the relevant unhatted ones $W_{m}^{\pm}$ in eqs. (\ref{ww1})-(\ref{ww4})  by making the re-scalings
\begin{eqnarray}
&& \widetilde{\psi}_{L}^{1} \rightarrow
 \,(\frac{\l_{1}}{2})^{1/4}
\widetilde{\psi}_{L}^{1},\,\,\,\, \widetilde{\psi}_{L}^{2} \rightarrow
 \,(\frac{\l_{2}}{2})^{1/4}
\widetilde{\psi}_{L}^{2},\,\,\,\,{\psi}_{L}^{3} \rightarrow
 \,(\frac{\l_{3}}{2})^{1/4}
{\psi}_{L}^{3}. \\
&& {\psi}_{L}^{1} \rightarrow
 \,(\frac{\delta_{1}}{2})^{1/4}
{\psi}_{L}^{1},\,\,\,\, {\psi}_{L}^{2} \rightarrow
 \,(\frac{\delta_{2}}{2})^{1/4}
{\psi}_{L}^{2},\,\,\,\,\widetilde{\psi}_{L}^{3} \rightarrow
 \,(\frac{\delta_{3}}{2})^{1/4}
\widetilde{\psi}_{L}^{3}. \\
&&  \widetilde{\psi}_{R}^{1} \rightarrow
 \,(\frac{\alpha_{1}}{2})^{1/4}
\widetilde{\psi}_{R}^{1},\,\,\,\, \widetilde{\psi}_{R}^{2} \rightarrow
 \,(\frac{\alpha_{2}}{2})^{1/4}
\widetilde{\psi}_{L}^{2},\,\,\,\,{\psi}_{R}^{3} \rightarrow
 \,(\frac{\alpha_{3}}{2})^{1/4}
{\psi}_{R}^{3}. \\
&& {\psi}_{R}^{1} \rightarrow
 \,(\frac{\beta_{1}}{2})^{1/4}
{\psi}_{R}^{1},\,\, {\psi}_{R}^{2} \rightarrow
 \,(\frac{\beta_{2}}{2})^{1/4}
{\psi}_{R}^{2},\,\,\,\,\widetilde{\psi}_{R}^{3} \rightarrow
 \,(\frac{\beta_{3}}{2})^{1/4}
\widetilde{\psi}_{R}^{3},
\end{eqnarray}
where the $\l_{j}\,\d_{j},\,\a_{j},\, \b_{j}$ are constant
parameters. These constants are introduced with the aim of
recovering some coupling constants between the currents of the
model.

Actually, in matrix form we have the following relationships $
\hat{W}_{m}^{+}=L_{m}^{+} W_{m}^{+} (L^{+}_{m})^{-1}$\, and\, $
\hat{W}_{m}^{-}=L^{-}_{m}W_{m}^{-} (L^{-}_{m})^{-1}$. The $L^{\pm}_{2}$, $L^{\mp}_{1}$ matrices, respectively,  take the following forms
\begin{eqnarray}
\left(
  \begin{array}{ccc}
    \sqrt[12]{\frac{x_{3}}{x_{1}}} & 0 & 0 \\
    0 &  \sqrt[12]{\frac{x_{1}}{x_{2}}} & 0 \\
    0 & 0 &  \sqrt[12]{\frac{x_{2}}{x_{3}}} \\
  \end{array}
\right)\;\; \mbox{and} \;\;  \left(
  \begin{array}{ccc}
    \sqrt[12]{\frac{y_{1}}{y_{3}}} & 0 & 0 \\
    0 &  \sqrt[12]{\frac{y_{2}}{y_{1}}} & 0 \\
    0 & 0 &  \sqrt[12]{\frac{y_{3}}{y_{2}}} \\
  \end{array}
\right),
\end{eqnarray}
supplied with the replacements $x \rightarrow \lambda$ for $L_{2}^{+}$, $y \rightarrow \beta$ for
$L_{2}^{-}$, $\;x \rightarrow
\alpha$ for $L_{1}^{-}$,  and \,$\;y \rightarrow \delta$ for $L_{1}^{+}$.

Some relationships between these parameters will emerge below  mainly
arising from the consideration of current-current (generalized
Thirring) type interactions among the various flavor species and
integrability requirement through the zero-curvature formulation
of the equations of motion.

In the following we will consider the eqs. of motion
(\ref{eqt11})-(\ref{eqt22}) in term of the field components. For future convenience let us introduce the fields $A_{R,\,L}^{i}$ as
\begin{eqnarray}
\label{ar1}
 A_{R}^{1} &=& \sqrt[4]{\frac{\alpha_{1} \beta_{1}}{4}} \,\psi^{1}_{R} \star
\tilde{\psi}^{1}_{R} +
 \sqrt[4]{\frac{\beta_{3} \alpha_{3}}{4}} \, \psi^{3}_{R} \star
 \tilde{\psi}^{3}_{R} \\\label{ar2}
A_{R}^{2} &=&  \sqrt[4]{\frac{\alpha_{2} \beta_{2}}{4}} \,
\psi^{2}_{R} \star \tilde{\psi}^{2}_{R} -\sqrt[4]{\frac{\alpha_{1}
\beta_{1}}{4}} \, \tilde{\psi}^{1}_{R} \star
 {\psi}^{1}_{R} \\\label{ar3}
 A_{R}^{3} &=&\sqrt[4]{\frac{\beta_{3} \alpha_{3}}{4}}\, \tilde{\psi}^{3}_{R} \star
{\psi}^{3}_{R} +
 \sqrt[4]{\frac{\alpha_{2} \beta_{2}}{4}}\, \tilde{\psi}^{2}_{R} \star
 {\psi}^{2}_{R}.
\end{eqnarray}
and
\begin{eqnarray}
\label{al1}
 A_{L}^{1} &=& \sqrt[4]{\frac{\delta_{1} \lambda_{1}}{4}} \,\psi^{1}_{L} \star
\tilde{\psi}^{1}_{L} +
 \sqrt[4]{\frac{\delta_{3} \lambda_{3}}{4}} \, \psi^{3}_{L} \star
 \tilde{\psi}^{3}_{L} \\
 \label{al2}
A_{L}^{2} &=&  \sqrt[4]{\frac{\delta_{2} \lambda_{2}}{4}} \,
\psi^{2}_{L} \star \tilde{\psi}^{2}_{L} -\sqrt[4]{\frac{\delta_{1}
\lambda_{1}}{4}} \, \tilde{\psi}^{1}_{L} \star
 {\psi}^{1}_{L} \\
 A_{L}^{3} &=&\sqrt[4]{\frac{\delta_{3} \lambda_{3}}{4}}\, \tilde{\psi}^{3}_{L} \star
{\psi}^{3}_{L} +
 \sqrt[4]{\frac{\delta_{2} \lambda_{2}}{4}}\, \tilde{\psi}^{2}_{L} \star
 {\psi}^{2}_{L}.\label{al3}
\end{eqnarray}

In terms of these fields the currents in (\ref{ncmt1}) become
\begin{eqnarray}
\hat{J}_{1}^{-}=\hat{J}_{2}^{-}=-\frac{i}{2}\left(
  \begin{array}{ccc}
    A_{R}^{1} & 0 & 0 \\
    0 &  A_{R}^{2} & 0 \\
    0 & 0 & -A_{R}^{3} \\
  \end{array}
\right)\;\; \mbox{and} \;\;  \hat{J}_{1}^{+}=\hat{J}_{2}^{+}=-\frac{i}{2}\left(
  \begin{array}{ccc}
    A_{L}^{1} & 0 & 0 \\
    0 &  A_{L}^{2} & 0 \\
    0 & 0 & -A_{L}^{3} \\
  \end{array}
\right)
\end{eqnarray}

Therefore the action of the $NCGMT_{1}$ model (\ref{ncmt1}) in terms of the Thirring field
components become
\begin{eqnarray} \label{ac1}
S_{NCGMT_{1}}&=& \int dx^2 \sum_{i=1}^{i=3} \Big\{\[2 i
\tilde{\psi}^{i}_{L} \star \partial_{+}{\psi}^{i}_{L} +
2i\tilde{\psi}^{i}_{R} \star
\partial_{-}{\psi}^{i}_{R} +i m_{i} (\tilde{\psi}^{i}_{L} \star
{\psi}^{i}_{R}
- \psi^{i}_{L} \star \tilde{\psi}^{i}_{R} )\] \nonumber \\
&- & 2(A_{L}^{i} \star A_{R}^{i}) \Big\},
\end{eqnarray}

Next let us write the equations of motion for the field components
derived from the action above. The following three equations of
motion
\begin{eqnarray}\label{eqs11}
\partial_{+} \psi^{3}_{L} &=& -\frac{1}{2} m_{3} \psi^{3}_{R} -
i \sqrt[4]{\frac{\delta_{3}\lambda_{3}}{4}} \,  \{ \psi^{3}_{L}
\star A^{3}_{R} + A^{1}_{R}
\star\psi^{3}_{L} \}\\
 \label{eqs12}
\partial_{+} \tilde{\psi}^{1}_{L} &=& -\frac{1}{2} m_{1}
\tilde{\psi}^{1}_{R} +  i
\sqrt[4]{\frac{\delta_{1}\lambda_{1}}{4}}\, \{
\tilde{\psi}^{1}_{L} \star A^{1}_{R} -
 A^{2}_{R} \star\tilde{\psi}^{1}_{L} \}\\
 \label{eqs13}
\partial_{+} \tilde{\psi}^{2}_{L} &=& - \frac{1}{2} m_{2}
\tilde{\psi}^{2}_{R} + i
\sqrt[4]{\frac{\delta_{2}\lambda_{2}}{4}}\, \{
\tilde{\psi}^{2}_{L} \star A^{2}_{R} + A^{3}_{R}
\star\tilde{\psi}^{2}_{L} \},
\end{eqnarray}
will correspond to the matrix form (\ref{eqt11}) for $m=1$.

One can obtain the equations of motion
\begin{eqnarray}\label{eqs21}
\partial_{+} \tilde{\psi}^{3}_{L} &=& -\frac{1}{2} m_{3} \tilde{\psi}^{3}_{R}
+ i  \sqrt[4]{\frac{\delta_{3}\lambda_{3}}{4}}\,  \{ A^{3}_{R}
\star
\tilde{\psi}^{3}_{L}  + \tilde{\psi}^{3}_{L} \star A^{1}_{R}  \}\\
 \label{eqs22}
\partial_{+} {\psi}^{1}_{L} &=& -\frac{1}{2} m_{1}
{\psi}^{1}_{R} -  i \sqrt[4]{\frac{\delta_{1}\lambda_{1}}{4}}\, \{
A^{1}_{R} \star {\psi}^{1}_{L}  - {\psi}^{1}_{L} \star A^{2}_{R}
 \}\\
 \label{eqs23}
\partial_{+} {\psi}^{2}_{L} &=& - \frac{1}{2} m_{2}
{\psi}^{2}_{R} - i \sqrt[4]{\frac{\delta_{2}\lambda_{2}}{4}} \{
A^{2}_{R} \star {\psi}^{2}_{L} + {\psi}^{2}_{L} \star A^{3}_{R}
\},
\end{eqnarray}
which in matrix form corresponds to eq. (\ref{eqt11}) for  $m=2$.

Similarly, one can obtain the equations of motion
\begin{eqnarray}\label{eq31}
\partial_{-} \psi^{3}_{R} &=& \frac{1}{2} m_{3} \psi^{3}_{L} - i
\sqrt[4]{\frac{\alpha_{3}\beta_{3}}{4}} \,
 \{ \psi^{3}_{R} \star A^{3}_{L} + A^{1}_{L} \star\psi^{3}_{R} \}
\\\label{eq32}
\partial_{-} \tilde{\psi}^{1}_{R} &=&  \frac{1}{2} m_{1}
\tilde{\psi}^{1}_{L} + i\sqrt[4]{\frac{\alpha_{1}\beta_{1}}{4}}
 \, \{ \tilde{\psi}^{1}_{R} \star A^{1}_{L}
 - A^{2}_{L} \star\tilde{\psi}^{1}_{R}
\}\\
\label{eq33}
\partial_{-} \tilde{\psi}^{2}_{R} &=& \frac{1}{2} m_{2}
\tilde{\psi}^{2}_{L}
 + i \sqrt[4]{\frac{\alpha_{2}\beta_{2}}{4}}
 \, \{\tilde{\psi}^{2}_{R} \star A^{2}_{L}
 + A^{3}_{L} \star\tilde{\psi}^{2}_{R} \},
\end{eqnarray}
corresponding to  $m=2$ in (\ref{eqt22}).

Finally, the equations
\begin{eqnarray}\label{eq41}
\partial_{-} \tilde{\psi}^{3}_{R} &=& \frac{1}{2} m_{3} \tilde{\psi}^{3}_{L} + i
\sqrt[4]{\frac{\alpha_{3}\beta_{3}}{4}}\,
 \{ A^{3}_{L} \star
\tilde{\psi}^{3}_{R}  + \tilde{\psi}^{3}_{R} \star A^{1}_{L} \}
\\\label{eq42}
\partial_{-} {\psi}^{1}_{R} &=&  \frac{1}{2} m_{1}
{\psi}^{1}_{L} - i \sqrt[4]{\frac{\alpha_{1}\beta_{1}}{4}}
 \, \{ A^{1}_{L} \star
{\psi}^{1}_{R}  - {\psi}^{1}_{R} \star A^{2}_{L}
\}\\
\label{eq43}
\partial_{-} {\psi}^{2}_{R} &=& \frac{1}{2} m_{2}
{\psi}^{2}_{L}
 - i \sqrt[4]{\frac{\alpha_{2}\beta_{2}}{4}}
 \, \{A^{2}_{L} \star
{\psi}^{2}_{R}  + {\psi}^{2}_{R} \star A^{3}_{L} \},
\end{eqnarray}
can be obtained from (\ref{eqt22}) in the case  $m=1$.

The set of  equations of motions (\ref{eqs11})-(\ref{eq43}) are the  $GL(3)$ extension
of the equations of motion given before for the case  $GL(2)$ NCMT$_{1}$ ( see eqs.
(5.11)-(5.14) of ref. \cite{jhep2}). In fact, the later system is contained in the $GL(3)$ extended model. For example, if one considers
$\psi^{1}_{L}=\psi^{2}_{L}=\tilde{\psi}^{1}_{L}=\tilde{\psi}^{1}_{L}=0$ in the eq. (\ref{eqs11}) then it is reproduced the equation (5.13) of reference
\cite{jhep2} describing the single Thirring field
$\psi_{3}$ provided that the parameters expression
$\sqrt[4]{\frac{\delta_{3}\lambda_{3}\beta_{3}\alpha_{3}}{16}} $
 corresponds to the coupling constant $\frac{\lambda}{2}$ of that reference.

The four field interaction terms in the action (\ref{ac1}) can be
re-written as a sum of Dirac type current-current terms for the
various flavors $(j=1,2,3)$.  In the constructions of the relevant
currents the double-gauging of a U(1) symmetry in the
star-localized Noether procedure deserves a careful treatment
\cite{liao, jhep2}. So, one has two types of currents for each
flavor \cite{jhep2} \br j_{k}^{(1)\,\mu}&=&\bar{\psi}_{k}
\gamma^{\mu} \star \psi_{k}, \label{j1}\\
j_{k}^{(2)\,\mu}&=&-\psi_{k}^{T}\g^{0}
\gamma^{\mu} \star \widetilde{\psi}_{k},\,\,\,\,\,\,\, k=1,2,3.\label{j2}.\er

Notice that in the commutative limit one has  $j_{k}^{(1)\,\mu}= j_{k}^{(2)\,\mu}$. In order to write as a sum of current-current interaction
terms it is necessary to impose the next constraints on the
$\a_{i},\b_{i}, \d_{i},\,\l_{i}$ parameters
\begin{eqnarray}
\label{kappa}
\frac{\delta_{j}\lambda_{j}}{\alpha_{j}\beta_{j}}= \kappa = \mbox{const.};\,\,\,\,\,\,j=1,2,3.
\end{eqnarray}

Then the four-spinor interactions terms in (\ref{ac1}), provided
that (\ref{kappa}) is taken into account,  can be written as
current-current interaction terms
\begin{eqnarray} \label{ncc}
-2 \sum_{i=1}^{3} A_{L}^{i} A_{R}^{i}&=& - g_{11} \, (j_{1\,\mu}^{(1)} \star j_{1}^{(1)\mu} +
j_{1\,\mu}^{(2)} \star j_{1}^{(2)\mu}) - g_{22} \,(j_{2\,\mu}^{(1)} \star
j_{2}^{(1)\mu} + j_{2\,\mu}^{(2)} \star j_{2}^{(2)\mu})  -\nonumber\\&&
g_{33}\,(j_{3\,\mu}^{(1} \star j_{3}^{(1)\mu} + j_{3\,\mu}^{(2)} \star
j_{3}^{(2)\mu} ) + g_{12}\,(j_{1\,\mu}^{(1)} \star
j_{2}^{(2)\mu})-\nonumber\\&&g_{23}\, ( j_{2\,\mu}^{(1)} \star j_{3}^{(1)\mu}) -
g_{13}(j_{1\,\mu}^{(2)} \star
j_{3}^{(2)\mu} ) \label{jj},
\end{eqnarray}
where
\begin{eqnarray}
g_{jj} = \frac{1}{4}  \sqrt[4]{\alpha_{j}\beta_{j}
\delta_{j} \lambda_{j}},\,\,\,\, g_{jk} = \frac{1}{2}  \sqrt[4]{\alpha_{j}\beta_{j}
\delta_{k} \lambda_{k}},\, (j\neq k);\;\,\,\,\,j,k=1,2,3.
\label{gij}
\end{eqnarray}

These parameters $g_{ij}$ define the coupling constants of the NC
generalized Thirring model (NCGMT$_{1}$), even though that they are not mutually
independent. Notice that considering the relationships
(\ref{kappa}) and  (\ref{gij}) one has the three constraints \br
g_{ij} =  2 \,\sqrt{g_{ii} \, g_{jj}}, \,\,\,\,i \neq
j.\label{constgij} \er

Taking into account the constraints (\ref{constgij}) we are left
with three  independent coupling parameters at our disposal, so in
order to study further properties such as the integrability and
the zero-curvature formulations of the model one must consider the
remaining three parameters, say the independent coupling parameters
$g_{11},\,g_{22},\,g_{33}$. Then, substituting in the action (\ref{ac1}) the current-current
 interaction terms (\ref{ncc}) one has
\begin{eqnarray}
S_{NCGMT_{1}}&=& \int dx^2 \Big\{\sum_{i=1}^{i=3}\[2 i
\tilde{\psi}^{i}_{L} \star \partial_{+}{\psi}^{i}_{L} +
2i\tilde{\psi}^{i}_{R} \star
\partial_{-}{\psi}^{i}_{R} +i m_{i} (\tilde{\psi}^{i}_{L} \star
{\psi}^{i}_{R}
- \psi^{i}_{L} \star \tilde{\psi}^{i}_{R} )\] \nonumber \\
&- & g_{11} \, (j_{1\,\mu}^{(1)} \star j_{1}^{(1)\mu} +
j_{1\,\mu}^{(2)} \star j_{1}^{(2)\mu}) - g_{22} \,(j_{2\,\mu}^{(1)} \star
j_{2}^{(1)\mu} + j_{2\,\mu}^{(2)} \star j_{2}^{(2)\mu})  -\nonumber\\&&
g_{33}\,(j_{3\,\mu}^{(1} \star j_{3}^{(1)\mu} + j_{3\,\mu}^{(2)} \star
j_{3}^{(2)\mu} ) + g_{12}\,(j_{1\,\mu}^{(1)} \star
j_{2}^{(2)\mu})-\nonumber\\&&g_{23}\, ( j_{2\,\mu}^{(1)} \star j_{3}^{(1)\mu}) -
g_{13}(j_{1\,\mu}^{(2)} \star
j_{3}^{(2)\mu} ) \Big\} \label{ncgmtcurr1}.
\end{eqnarray}

We define this model as the NC (generalized) massive Thirring model NCGMT$_{1}$ written in terms of the component fields. Its matrix version is understood to be the action (\ref{ncmt1}) once the parameters relationships (\ref{kappa}) are taken into account.

The two types of U(1) currents $j_{k\,\mu}^{(1)},\,
j_{k\,\mu}^{(2)}$ (k=1,2,3), respectively, satisfy the conservation equations
\begin{eqnarray}
 \partial_{+} (\tilde{\psi}^{k}_{L} \star
{\psi}^{k}_{L})  + \partial_{-}( \tilde{\psi}^{k}_{R} \star
{\psi}^{k}_{R})  =0,\,\,\,\,\,  \partial_{+} (\psi^{k}_{L} \star
\tilde{\psi}^{k}_{L})  +
   \partial_{-}( \psi^{k}_{R} \star \tilde{\psi}^{k}_{R}) =0,\,\,\,\,\,k=1,2,3
   \label{currents}.
\end{eqnarray}

\subsubsection{(Constrained) NC(c)GMT$_{1}$ zero-curvature formulation}
\label{zerocurvature}

The zero-curvature condition encodes integrability even in the NC
extension of integrable models (see e.g. \cite{jhep2}  and
references therein), as this condition allows, for example, the
construction of infinite conserved charges for them. In order to
tackle this problem it is convenient to consider the matrix form
of the equations of motion of the $GL(3)$ NC Thirring model
(\ref{eqt11})-(\ref{eqt22}) and intend to write them as
originating from a zero-curvature condition. So, taking into
account the gradation structure of the model let us consider the
following Lax pair
\begin{eqnarray} \nonumber
A_{-}&=&E_{-3} +a[E_{-3},W^{+}_{1}]_{\star} +
b[E_{-3},W^{+}_{2}]_{\star}+ g_{1}
[[E_{-3},\hat{W}^{+}_{1}],\hat{W}^{+}_{2}]_{\star} + g_{2}
[[E_{-3},\hat{W}^{+}_{2}],\hat{W}^{+}_{1}]_{\star}. \\\label{laxp1} \\
A_{+}&=&-E_{+3} + b[E_{+3},W^{-}_{1}]_{\star} +
a [E_{+3},W^{-}_{2}]_{\star} +\widetilde{g}_{1}
[[E_{+3},\hat{W}^{-}_{1}],\hat{W}^{-}_{2}]_{\star}
+\widetilde{g}_{2}
[[E_{+3},\hat{W}^{-}_{2}],\hat{W}^{-}_{1}]_{\star}, \nonumber \\
\label{laxp2}
\end{eqnarray}
where $a, b, g_{1}, g_{2}, \widetilde{g}_{1}, \widetilde{g}_{2}$
are some parameters to be determined below. Notice that the
potentials  $A_{\pm}$ lie in the directions of the  affine Lie
algebra generators of grade ${\cal G}_{0,1,2,3}$ and ${\cal
G}_{0,-1,-2,-3}$, respectively.

These matrix valued fields must be replaced into the
zero-curvature equation
\begin{eqnarray}\label{curvnula}
\[ \partial_{+} +  A_{+}\,, \, \partial_{-} +  A_{-} \]_{\star} = 0,
\end{eqnarray}
We will use the following relationships which can easily be established
\br
4\hat{J}^{-}_{1}&=&4\hat{J}^{-}_{2}=
-[[E_{+3},\hat{W}^{-}_{1}],\hat{W}^{-}_{2}]_{\star}=-
[[E_{+3},\hat{W}^{-}_{2}],\hat{W}^{-}_{1}]_{\star},\label{j11}\\
4\hat{J}^{+}_{1}&=&4\hat{J}^{+}_{2}=
[[E_{-3},\hat{W}^{+}_{1}],\hat{W}^{+}_{2}]_{\star}=
[[E_{-3},\hat{W}^{+}_{2}],\hat{W}^{+}_{1}]_{\star}\label{j22}\er

So, the Lax pair  can be rewritten as
\begin{eqnarray} \label{laxp11}
A_{-}&=&E_{-3} +a[E_{-3},W^{+}_{1}]_{\star} +
b[E_{-3},W^{+}_{2}]_{\star}+ k_1 \hat{J}^{+}_{1}. \\
A_{+}&=&-E_{+3} + b[E_{+3},W^{-}_{1}]_{\star} + a
[E_{+3},W^{-}_{2}]_{\star} + k_2 \hat{J}^{-}_{1}, \label{laxp22}
\end{eqnarray}
where  we have introduced the new parameters $k_{1,\,2}$ such that
$\widetilde{g}_{1}+\widetilde{g}_{2}=-\frac{k_2}{4}$, and ${g}_{1}
+ {g}_{2}=\frac{k_1}{4}$

In order to get the relevant equations of motion
(\ref{eqt11})-(\ref{eqt22}) it is useful to take into
consideration the gradation structure of the various terms. So,
the terms of gradation $(-1)$ in (\ref{curvnula}), taking into
account (\ref{j11}), become
\begin{eqnarray}
\[ E_{-3}, \partial_{+}{W}^{+}_{2} \]_{\star} &= & + [E_{-3},
[E_{3},{W}^{-}_{1}]]_{\star} - k_2 (L_{2}^{+})^{-1}
[\hat{J}^{-}_{1}, [E_{-3},\hat{W}^{+}_{2}]]_{\star}L^{+}_{2} + \[
F_{1}^{+}, F_{2}^{-}
\]_{\star},\nonumber \\\label{cc1}
\end{eqnarray}

The equation (\ref{cc1})  has the same structure as the equation
of motion (\ref{eqt11}) ( for $m=1$) provided that we set
$L_{2}^{+} = L_{1}^{+}$\, , and impose the constraint
\begin{eqnarray}\label{cons1}
 \[ F_{1}^{+},  F_{2}^{-}
\]_{\star}=0.
\end{eqnarray}

Next, looking for the gradation $(+1)$  terms in (\ref{curvnula}) and using (\ref{j22}) we  may get the equation
\begin{eqnarray}\label{cc2}
 \[ E_{3}, \partial_{-} W^{-}_{2} \]_{\star} & = & -
[E_{3},[E_{-3},{W}^{+}_{1}]]_{\star}  -k_1 (L_{2}^{-})^{-1}
[\hat{J}^{+}_{1}, [E_{3},\hat{W}^{-}_{2}]]_{\star} L^{-}_{2} + \[
F_{2}^{+},  F_{1}^{-} \]_{\star}.
\end{eqnarray}

In a similar way, identifying  $L_{2}^{-} = L_{1}^{-}$, and
imposing the constraint
\begin{eqnarray}\label{cons2}
\[ F_{2}^{+},  F_{1}^{-} \]_{\star}=0,
\end{eqnarray}
one notices that the equation (\ref{cc2}) is equal to the
equation of motion (\ref{eqt22}) (for $m=1$).

Following the process we can write for the $(\pm 2)$ gradations and conclude
that in order to obtain the two equations of motion in
(\ref{eqt11})-(\ref{eqt22}) \, for $m= 2$,
 it is required the same conditions $L_{2}^{\pm} = L_{1}^{\pm}$ as above,
 without any new constraint.

 We notice that the conditions $L_{2}^{\pm} = L_{1}^{\pm}$ which are related to the equations of motion
 for the gradations $(\pm 1), (\pm 2)$ provide the following constraints between the initial parameters ($\a_{i},\b_{i}, \l_{i}, \d_{i}$)
\br
\label{const2}
\a_{i}\b_{i}=r_{1};\,\,\,\,\,\l_{i}\d_{i}=r_{2},\,\,\,\,i=1,2,3;\,\,\,\,r_{1},\,\,r_{2}=\mbox{constants.}
\er

In fact, these constraints are consistent with  the parameters relationships
(\ref{kappa}) established   above; however, eqs. (\ref{const2})
incorporate additional constant parameters $r_1,\, r_2 $ such that  $\kappa = r_{2}/r_{1}$.
Additional relationships between the parameters arise by requiring
that the above matrix equations derived from the zero-curvature
equation to be consistent with the eqs. of motion
(\ref{eqs11})-(\ref{eq43}). So, together with the relationships
(\ref{const2}), it is required
\br \nonumber \a_1 \a_2\a_3=\b_1 \b_2\b_3\equiv r_{1}^{3/2};\,\,\,\,\, \l_1
\l_2\l_3=\d_1 \d_2\d_3\equiv r_2^{3/2},\,\,\,k_{1} &=& 2^{3/4}
r_1^{1/8},\,\,\,\,\, k_{2} = 2^{3/4} r_2^{1/8}\\\label{alfasbetas}\er

So, the set of current-current coupling constants $g_{ij}$ in (\ref{ncgmtcurr1}), which in the last section have been  assumed to be equivalent to three independent parameters, in view of the  additional relationships (\ref{alfasbetas})   they
reduce to only one independent parameter $g$ defined by
 \br g_{12}&=&g_{23}=g_{13}=\frac{1}{2} g; \,\,\,\,\,\,
g_{ii}=\frac{1}{4} g, \,\,\,i=1,2,3;\,\,\,g\equiv (r_{1}
r_{2})^{1/4}. \er

Finally, for the zero gradation term there appears the following equation
\begin{eqnarray}\label{c3zero}
&& k_1 \partial_{+} \hat{J}_{1}^{+} -
k_2 \partial_{-} \hat{J}_{1}^{-}  - a b [F_{2}^{+}
,F_{2}^{-}]  - a b[F_{1}^{+},F_{1}^{-}] + k_1 k_2
[\hat{J}_{1}^{-} ,\hat{J}_{1}^{+}] =0.
\end{eqnarray}

We require this equation to be consistent with the full equations of motion (\ref{eqs11})-(\ref{eq43}) and the constraints (\ref{cons1}) and (\ref{cons2}). These constraints in terms  of the fundamental fields become
\begin{eqnarray}
\label{cons11}
\psi^{1}_{R} * \psi^{2}_{L} = \psi^{1}_{L} * \psi^{2}_{R}, \;\,\,
\psi^{2}_{R}* \tilde{\psi}^{3}_{L} = -\psi^{2}_{L} *
\tilde{\psi}^{3}_{R},\;\,\, \tilde{\psi}^{3}_{L} * \psi^{1}_{R}=-
\tilde{\psi}^{3}_{R} * \psi^{1}_{L}
\end{eqnarray}
and
\begin{eqnarray}
\psi^{3}_{R} * \tilde{\psi}^{2}_{L} = -\psi^{3}_{L} *
\tilde{\psi}^{2}_{R}, \;\,\,\tilde{\psi}^{1}_{L}* {\psi}^{3}_{R} =-
\tilde{\psi}^{1}_{R} * {\psi}^{3}_{L},\;\,\, \tilde{\psi}^{2}_{R}
* \tilde{\psi}^{1}_{L}= \tilde{\psi}^{2}_{L} * \tilde{\psi}^{1}_{R},
\label{cons22}\end{eqnarray}
respectively.

In order to establish specific relationships between the parameters $a,b$ and $r_1, r_2$ let us write  (\ref{c3zero}) in terms of the fundamental fields
\begin{eqnarray}
i (k_1\partial_{+} A^{1}_{L} -k_2
\partial_{-} A^{1}_{R} ) &=&-\frac{k_1 k_2}{2}(A^1_{R}\star A^1_{L}- A^1_{L}\star A^1_{R})-2 a b \{ i m_{1} (
\sqrt[4]{\frac{\beta_{1}\lambda_{1}}{4}}{\psi}^{1}_{R} \star
\tilde{\psi}^{1}_{L}+\nonumber\\
&&\sqrt[4]{\frac{\alpha_{1}\delta_{1}}{4}}{\psi}^{1}_{L} \star
\tilde{\psi}^{1}_{R} )+ 
i m_{3} (
\sqrt[4]{\frac{\alpha_{3}\delta_{3}}{4}}{\psi}^{3}_{R} \star
\tilde{\psi}^{3}_{L}+
\sqrt[4]{\frac{\beta_{3}\lambda_{3}}{4}}{\psi}^{3}_{L} \star
\tilde{\psi}^{3}_{R} ) \}\nonumber\\ 
\label{zero1}\\
i (k_1 \partial_{+} A^{2}_{L} -k_2
\partial_{-} A^{2}_{R} )_{\star} &=& -\frac{k_1 k_2}{2}( A^{2}_{R} \star A^{2}_{L}-
A^{2}_{L} \star A^{2}_{R}) -2 a b \{i m_{2} (
\sqrt[4]{\frac{\beta_{2}\lambda_{2}}{4}}{\psi}^{2}_{R} \star
\tilde{\psi}^{2}_{L}+\nonumber\\&&
\sqrt[4]{\frac{\alpha_{2}\delta_{2}}{4}}{\psi}^{2}_{L} \star
\tilde{\psi}^{2}_{R}  )-  i m_{1}
(\sqrt[4]{\frac{\delta_{2}\alpha_{2}}{4}} \tilde{\psi}^{1}_{R}
\star {\psi}^{1}_{L} +
\sqrt[4]{\frac{\beta_{1}\lambda_{1}}{4}}\tilde{\psi}^{1}_{L}
\star {\psi}^{1}_{R} ) \} \nonumber \\ 
\label{zero2} \\
 i(k_1 \partial_{+} A^{3}_{L} - k_2
\partial_{-} A^{3}_{R} )_{\star} &=& \frac{k_1 k_2}{2}( A^{3}_{R} \star A^{3}_{L}- A^{3}_{L} \star A^{3}_{R})   - 2 a b \{i m_{3} (
\sqrt[4]{\frac{\beta_{3}\lambda_{3}}{4}}\tilde{\psi}^{3}_{R} \star
{\psi}^{3}_{L}+ \nonumber \\&&
\sqrt[4]{\frac{\delta_{3}\alpha_{3}}{4}}
\tilde{\psi}^{3}_{L} \star {\psi}^{3}_{R}  )+ i m_{2} (
\sqrt[4]{\frac{\delta_{2}\alpha_{2}}{4}} \tilde{\psi}^{2}_{R}
\star {\psi}^{2}_{L}+ \sqrt[4]{\frac{\beta_{2}\lambda_{2}}{4}}
\tilde{\psi}^{2}_{L} \star
{\psi}^{2}_{R} ) \}. \nonumber \\
\label{zero3}
\end{eqnarray}

Substituting the fields $A_{R,L}^{j}, \,j=1,2,3$ in the form (\ref{ar1})-(\ref{al3}) into the eqs. (\ref{zero1})-(\ref{zero3}) and taking into account the set of equations of motion (\ref{eqs11})-(\ref{eq43}) one gets the following relationships
\br
2 ab = \sqrt{\frac{g}{2}}\,;\,\,\,\,\,(2)^{1/4}=r_{1}^{1/8}+r_{2}^{1/8}.
\er

Therefore, we have established a zero-curvature formulation of a constrained version of the NCGMT$_{1}$ model. From this point forward this constrained model will be dubbed as NC(c)GMT$_{1}$. 
 
Notice that the set of equations (\ref{zero1})-(\ref{zero3}) contain the relevant eq. associated to the $SL(2)$ NC massive Thirring model written for its relevant
zero gradation sector analogous to (\ref{c3zero}). So, for example, if one reduces the eq. (\ref{zero3}) to get an equation for just one field, say $\psi^{3}$, one has
\begin{eqnarray}
i \[k_1 \sqrt[4]{\frac{r_2}{4}} \partial_{+}
(\tilde{\psi}^{3}_{L}\star {\psi}^{3}_{L}) - k_2
\sqrt[4]{\frac{r_1}{4}} \partial_{-}
(\tilde{\psi}^{3}_{R}\star {\psi}^{3}_{R})\] &=&  -2 i a b m_{3} \(
\sqrt[4]{\frac{\beta_{3}\lambda_{3}}{4}}\tilde{\psi}^{3}_{R} \star
{\psi}^{3}_{L}+ \nonumber \\
&& \sqrt[4]{\frac{\delta_{3}\alpha_{3}}{4}}
\tilde{\psi}^{3}_{L} \star {\psi}^{3}_{R}  \)  + \frac{k_1 k_2}{2}\sqrt[4]{\frac{r_1 r_2 }{16}}
(\tilde{\psi}^{3}_{R}\star {\psi}^{3}_{R}\star\nonumber \\&&
\tilde{\psi}^{3}_{L}\star {\psi}^{3}_{L}-
\tilde{\psi}^{3}_{L}\star {\psi}^{3}_{L} \star
\tilde{\psi}^{3}_{R}\star {\psi}^{3}_{R}).\nonumber\\
\label{psi3eqcons}
\end{eqnarray}

Now, taking into account $\a_{3}=\b_{3}=\d_{3}=\l_{3}$
[$r_1= r_2\equiv r$] and the identifications $\psi^{3} \rightarrow  i\, r^{1/16}\psi$,\,\,\,
$r^{1/2} \rightarrow \lambda$, \, $
[m_{3} \frac{r^{1/8}}{2^{5/4}}] \rightarrow m_{\psi}$\, we arrive at the equation $
\partial_{-}
(\tilde{\psi}_{R}\star {\psi}_{R})-
 \partial_{+}
(\tilde{\psi}_{L}\star {\psi}_{L})  =  m_{\psi} \(
\tilde{\psi}_{R} \star {\psi}_{L}+ \tilde{\psi}_{L}
\star {\psi}_{R}  \) -i\l(\tilde{\psi}_{R}\star
{\psi}_{R}\star \tilde{\psi}_{L}\star {\psi}_{L}-
\tilde{\psi}_{L}\star {\psi}_{L} \star
\tilde{\psi}_{R}\star {\psi}_{R})$, which is
the eq. (5.18) of the ref. \cite{jhep2} .

\subsubsection{NCGMT$_1$ sub-models}
\label{NCGMTsubmodels}

In the following we discuss some reduced models associated to the
action (\ref{ncgmtcurr1}) and its equations of motion
(\ref{eqs11})-(\ref{eq43}).

{\bf NC massive Thirring (NCMT$_1$) models}

The reduction of the NCGMT$_{1}$ model equations  of motion
(\ref{eqs11})-(\ref{eq43}) to a model with just one spinor field, say
the components $\psi_{R,L}^{1},\, \widetilde{\psi}_{R,L}^{1}$ (consider the
reduction $\psi^{2, 3}_{R,\,L}=\widetilde{\psi}^{2, 3}_{R,\,L}=0$)
reproduces the NCMT$_{1}$ model which has been presented in
\cite{jhep2, jhep3}. Notice that in this case the constraints
(\ref{cons11}) and (\ref{cons22}), as well as  the decoupling
equations (\ref{sumconst1}) [or in components
(\ref{conscomp1})-(\ref{conscomp22})] become trivial. Let us emphasize that the
full decoupling eqs. are satisfied  by a subset of soliton solutions of the
field equations of the $GL(2)$ NCATM$_{1}$ model such that the two
sectors NCSG$_{1}$/NCMT$_{1}$ completely decouple \cite{jhep2}. Reducing in
this way it is clear the appearance of three copies of the
NCMT$_{1}$ model associated to the spinors $\psi^{1},\,
\psi^2$\,and\, $\psi^3$, respectively.

{\bf NC Bukhvostov-Lipatov (NCBL$_1$) model}

Consider a reduced model with two fields, say $\psi_{R,L}^{1,
2},\, \widetilde{\psi}_{R,L}^{1,2}$,  achieved through the
reduction $\psi^{ 3}_{R,\,L}=\widetilde{\psi}^{3}_{R,\,L}=0$.  So,
the Lagrangian (\ref{ncgmtcurr1}) becomes
\begin{eqnarray} \label{ncblm}
S_{NCTM}&=& \int dx^2 \Big\{ \sum_{i=1}^{i=2} \[2 i
\tilde{\psi}^{i}_{L} \star \partial_{+}{\psi}^{i}_{L} +
2i\tilde{\psi}^{i}_{R} \star
\partial_{-}{\psi}^{i}_{R} +i m_{i} (\tilde{\psi}^{i}_{L} \star
{\psi}^{i}_{R}
- \psi^{i}_{L} \star \tilde{\psi}^{i}_{R} )\] \nonumber \\
 && - g_{11} \, (j_{1\,\mu}^{(1)} \star j_{1}^{(1)\mu} +
j_{1\,\mu}^{(2)} \star j_{1}^{(2)\mu}) - g_{22}
\,(j_{2\,\mu}^{(1)} \star j_{2}^{(1)\mu} + j_{2\,\mu}^{(2)} \star
j_{2}^{(2)\mu})\nonumber\\ &&+ g_{12}\,(j_{1\,\mu}^{(1)} \star
j_{2}^{(2)\mu})\Big\}.
\end{eqnarray}

Remember that in ordinary space there is no distinction between
the type of $j^{(1)}_{i}$ and $j^{(2)}_{i}$ currents for each
flavor $i$ ; so, the model (\ref{ncblm}) when written in ordinary
space-time is known in the literature as the Bukhvostov-Lipatov
model (BL) \cite{lipatov}. It has been claimed the classical
integrability of the model in two special cases $g_{12}=0$
(2$\times$ MT model) and $g_{11}=g_{22}=0$ (BL model) [in both
cases consider $m_1=m_2$](see \cite{ameduri} and refs. therein).
The quantum integrability of the BL model has been discussed in
\cite{saleur}. In view of the above discussion we define the model
(\ref{ncblm}) as the first version of the NC
Bukhvostov-Lipatov model (NCBL$_{1}$). Actually, there are additionally 
 two reduction processes to arrive at NCBL$_{1}$ models, i.e. by
 setting $\psi^{1}=0$ and $\psi^{2}=0$ in (\ref{ncgmtcurr1}), respectively.

{\bf (Constrained) NC Bukhvostov-Lipatov (NC(c)BL$_1$) and Lax pair formulation}

Let us discuss a constrained version of the model (\ref{ncblm}). In view of the developments above one can establish the
zero-curvature formulation of a constrained model  associated to the model (\ref{ncblm}) by setting
$\psi_{L.R}^{3}=\widetilde{\psi}_{L.R}^{3}=0$ in the matrices
$W_{1,2}^{\pm}$ of the Lax pair  eqs. (\ref{laxp1})-(\ref{laxp2}),
provided  the constraints (\ref{cons11}) and (\ref{cons22}) given
in the form $\psi^{1}_{R}
* \psi^{2}_{L} = \psi^{1}_{L} * \psi^{2}_{R}$  and
$\tilde{\psi}^{2}_{R}
* \tilde{\psi}^{1}_{L}= \tilde{\psi}^{2}_{L} *
\tilde{\psi}^{1}_{R}$, are considered. So, we claim that the model
(\ref{ncblm}) is classically integrable provided that the above
constraints are taken into account. In this way, provided that for version 2 one writes a copy of the model and their relevant constraints, one defines the (constrained) NC(c)BL$_{1,2}$ models amenable to a Lax pair formulation . 

In connection to this
development, let us mention that  a version of the BL model for Grassmanian fields in
usual space-time has also been recently shown to be associated to
a Lax pair formulation provided some constraints are imposed
\cite{zimerman}.

In Fig. 1 we have outlined the various relationships. Notice that we have the two versions of NCATM$_{1,\,2}$ and their strong/weak sectors described by the models NCGSG$_{1,\,2}$ and  NCGMT$_{1,\,2}$, respectively, as well as the relevant sub-models. We have emphasized the field contents in each stage of the reductions. 

Some comments are in order here.

1. The action (\ref{ncgmtcurr1}) (or its matrix form
(\ref{ncmt1})) defines a three species NC generalized massive
Thirring model. We have tried to write its eqs. of motion
(\ref{eqt11})-(\ref{eqt22}) [or in components
(\ref{eqs11})-(\ref{eq43})] as deriving from a zero-curvature
formulation. We have proposed a Lax pair reproducing the same
equations of motion provided that the constraints (\ref{cons1})
and (\ref{cons2})[or in components (\ref{cons11}) and
(\ref{cons22})] are imposed. This fact suggests that the
NCGMT$_{1}$ model (\ref{ncmt1}) becomes integrable only for a
sub-model defined by the eqs. of motion (\ref{eqs11})-(\ref{eq43})
provided the constraints (\ref{cons1}) and (\ref{cons2}) are
satisfied \cite{posleny}. So, one expects that a careful
introduction of the constraints trough certain Lagrange
multipliers into the action will provide the Lagrangian
formulation of an integrable sub-model of the NCGMT$_{1}$ theory.

2. Regarding the action related  to  the full zero-curvature
equations of motion without constraints, determined by the set of
eqs. (\ref{cc1}) and (\ref{cc2}), and the relevant eqs. in
(\ref{eqt11})-(\ref{eqt22}) written for $m = 2$, it
is interesting to notice that the quadratic terms in the spinors
present in the first couple of eqs. of motion (\ref{cc1}) and 
(\ref{cc2}) make it difficult to believe that one can find a local
Lagrangian for the theory. Obviously, in that case we could not
have a generalized massive Thirring model with a local Lagrangian
involving bilinear (kinetic and mass terms) and usual
current-current terms. This fact is intimately related to the
presence of the eqs. (\ref{sumconst1}) [or in components
(\ref{conscomp1})-(\ref{conscomp22})] in the set of decoupling
eqs. (\ref{map1})-(\ref{currpm}). In the commutative case the
equations of type (\ref{sumconst1}) have been incorporated in
order to write a local Lagrangian for the GATM model in ref.
\cite{jhep1}. Notice that the original theory (without constraints)
allows a zero-curvature formulation; in fact, its Lax pair is just
the one of the so-called conformal affine Toda model coupled to
matter fields \cite{matter}. However, it does not posses a local Lagrangian
formulation in terms of the fields of the model; namely, the Toda and the spinor (Dirac) fields.

3. Notice that in Fig. 1  we have emphasized the duality relationship NCGSG$_1$ $\leftrightarrow$ NCGMT$_1$ since in this case the symmetry $U(1) \times U(1)\times U(1)$  of the NCGSG$_1$ model is implemented in the star-localized Noether procedure to get the three $U(1)$ currents of the NCGMT$_1$ sector. Regarding the relationships between the sub-models of the both sectors NCGSG$_1$ and NCGMT$_1$, it is clear the appearance of the duality NCSG$_1$ $\leftrightarrow$ NCMT$_1$ which has been discussed in the literature \cite{jhep2, jhep3}. This duality has also been discussed in the context of noncommutative bosonization of the massive Thirring model \cite{schiappa}. In addition, it is expected the duality relationship  NCbBL$_1$ $\leftrightarrow$ NCBL$_1$, since in the ordinary space-time the former is the bosonized version of the later model \cite{ameduri, saleur}. Regarding this type of duality relationships between the remaining models a more careful investigation is needed, e.g. we have not been able to describe neither the spinor model corresponding to the NCDSG$_{1}$ model, nor the scalar sectors of the (constrained) NC(c)GMT$_{1}$ and  NC(c)BL$_{1}$ models, respectively.  

\subsection{NCGMT$_{2}$}
\label{bncmt2}

As mentioned in the last paragraph of section \ref{decoupled} we expect
that another NCGMT$_{2}$ version, with twice the number of fields of the  NCGMT$_{1}$ theory, will appear when one performs  a similar
decoupling procedure for the extended system with $\{F^{\pm}_{m},
W^{\pm}_{m}\}$ and $\{{\cal F}^{\pm}_{m}, {\cal W}^{\pm}_{m}\}$ fields. In fact, a
copy of the NCGMT$_{1}$ action (\ref{ncmt1}), as well as the relevant
zero-curvature equation of motion can be written for the fields
$\{{\cal F}^{\pm}_{m}, {\cal W}^{\pm}_{m}\}$. Following similar steps one can construct a copy for each one of the sub-models presented above. Since it involves a direct generalization we will not present more details; however, see a corresponding construction for the $GL(2)$ case in ref. \cite{jhep2}. In this way one can get the NCGMT$_{2}$ model which is expected to be related to the NCGSG$_2$ model. Similarly to the NCGMT$_{1}$ case, one can expect that only a sub-model of NCGMT$_{2}$ will posses a zero-curvature formulation provided that a set of constraints similar to the eqs. (\ref{cons1}) and (\ref{cons2}), and a copy of them written for the fields ${\cal F}^{\pm}_{m}\,\,(m=1,2)$ are considered.

\section{Non-commutative solitons and kinks}
\label{ncsol}

It is a well known fact that the one-soliton solutions of certain models solve
their NC counterparts. This feature holds for the SG model and its NCSG$_{1,2}$ counterparts \cite{jhep2}. In the multi-field models, this feature means that the GSG model and its NCGSG$_{1, 2}$ extensions
 have a common subset of solutions, in particular the one-soliton and kink type solutions as we will see below. Of course the additional
constraints, in the form of conservation laws which we have described before, e.g. the eqs. (\ref{ncdsg12}) and (\ref{ncneweq22})-(\ref{ncneweq23}), respectively in the two versions of NCDSG models, must also be verified for the common subset of solutions. In fact, as we have noticed before they become trivial equations in the commutative limit.

The properties mentioned above reside on a simple observation: it is known that if $f(x_{0}, x_{1})$ and $g(x_{0}, x_{1})$
depend only on the combination $(x_{1}-v x_{0})$, then the product $f\star g$
coincides with the ordinary product $f.g$~\cite{cabrera, Dimakis}.
Therefore, all the $\star$ products in the NCGSG$_{1}$ system (\ref{ncsgeq21})-(\ref{ncsgeq3}) reduce to the ordinary ones, so for these types of functions one has:  NCGSG$_{1}$ $\rightarrow$ GSG model; the GSG model was defined in (\ref{limit111})-(\ref{limit222}) [see also  eqs. (\ref{sys1})-(\ref{sys2})]. In the following we record the solutions with this property, i.e, the one-soliton solutions of the NCGSG$_{1}$ model and the kink type solution of the NCDSG$_{1}$ sub-model. Actually, the same analysis can be done for the NCGSG$_{2}$ case.

\subsection{Solitons and kinks}
\label{s11}

Next we write  the 1-soliton and 1-kink type solutions
associated to the fields $\phi_{1, 2}$ of the NCGSG$_{1}$ model, which in accordance to the discussion above reduce to the GSG system of eqs. (\ref{limit111})-(\ref{limit222}). We will see that these solitons are, in fact, associated to the various sine-Gordon models obtained as sub-models of the GSG theory, and the kink type solution corresponds to the double sine-Gordon sub-model \cite{jhep4}.

1. Taking  $\phi_{1}=-\phi_{2}$ and $M_{3}=M_{2},\,\,\d_{i}=0$ in (\ref{limit111})-(\ref{limit222}) one has
\br \label{sol1}
\phi_{1}= 4 \mbox{arctan}\{d\,\, \mbox{exp}[\gamma_{1}
(x-v t)]\}. \er

2. For  $\phi_{1}=\phi_{2}$ and $M_{2}=M_{3},\,\,M_{1}=0$ one has
\br \label{sol12}
\phi_{1}= 4\, \mbox{arctan}\{d\,\, \mbox{exp}[\gamma_{2}
(x-v t)]\}. \er

Another SG model is given by setting $\phi_{1}=\phi_{2}$ and $M_{2}=M_{3}=0$ in (\ref{limit111})-(\ref{limit222}) which leads to another soliton solution.

3. The kink solution is associated to the reduced double sine-Gordon model obtained by taking   $\phi_{1}=\phi_{2}\equiv \phi$ and $M_{3}=M_{2},\,\,M_{j} \neq 0$. So, one has

\br \label{kink} \phi
:=4 \, \mbox{arctan}\left[ d_{K} \,\,\mbox{sinh}[\gamma_{K}\, (x-vt) ]\right], \er
which is the usual DSG kink solution \cite{campbell}.

The $\g_{1,2}, \g_{K}, d , d_{K}, v$ above are some constant parameters.

\section{Conclusions and discussions}
\label{concl}

Some properties of the NC extensions of the GATM model and their
weak-strong phases described by the NCGMT$_{1,2}$ and NCGSG$_{1,2}$
models, respectively, have been considered. The Fig. 1 summarizes
the relationships we have established, as well as the field contents in each sub-model.

In the
$\theta \rightarrow 0$ limit we have the following correspondences: NCATM$_{1, 2}
\rightarrow$ GATM;\, NCGSG$_{1,2}$ (the real sector of model $2$)
$\rightarrow$ GSG(plus a free scalar in the case of model $2$); NCbBL$_{1,2}$ $\rightarrow$ bBL; NCDSG$_{1,2}$ (the real sector of model $2$) $\rightarrow$ DSG(plus a free scalar in the case of model $2$);
 NCGMT$_{1, 2}$ $\rightarrow$ GMT (two copies in case of model 2); NCBL$_{1, 2}$ $\rightarrow$ BL (two copies in case of model $2$). In addition, the constrained versions  NC(c)GMT$_{1, 2}$ and NC(c)BL$_{1, 2}$ give rise, in this limit,  to the relevant (constrained) GMT and BL models, respectively,  in ordinary space. To our knowledge, these are novel spinor integrable models.   

The NCGMT$_{1,2}$ Lagrangians describe
three flavor massive spinors (case 2 considers twice the number of spinors) with current-current interactions
among themselves. In the process of constructing the Noether currents one recognizes
the $[U(1)]^3$ symmetry in both NCGMT$_{1, 2}$ models (in fact, as
a subgroup of $[U(1)_{C}]^3$ in the model 2).
We have provided the zero-curvature formulation of certain sub-models of the NCGMT$_{1,2}$. In fact, in order to write the eqs. of motion (\ref{eqs11})-(\ref{eq43}) as a  zero-curvature  equation for a suitable Lax pair one needs to impose the constraints (\ref{cons11})-(\ref{cons22}), defining in this way the NC(c)GMT$_{1,2}$ models. Likewise, the (constrained) NC(c)BL$_{1,2}$ models possess certain Lax pairs.    

The generalized sine-Gordon model, the usual SG model, the Bukhvostov-Lipatov model and the double sine-Gordon theory appear in the
commutative limit of the both versions of the NCGSG$_{1,2}$ models.
We have concluded that the NCGSG$_{1, 2}$ models possess the same soliton and kink type solutions as their commutative counterparts. The appearance
of the non-integrable double sine-Gordon model as a sub-model of
the GSG model suggests that even the NCGSG$_{1,2}$ models are non-integrable theories
for the arbitrary set of values of the parameter space, since they possess as sub-models the corresponding NCDSG$_{1,2}$ models. However, the NCGSG$_{1, 2}$ models possess certain integrable directions in field space, as remarkable examples one has the NCSG$_{1,2}$ sub-models. In view of the presence of the  (constrained) NC(c)GMT$_{1, 2}$ and NC(c)BL$_{1, 2}$ models with corresponding zero-curvature formulations, it is expected  the existence of other integrable directions in the scalar sector, which we have not pursued further in the present work.

Actually, the procedures presented so far can directly be extended to the NCATM model for the affine
Lie algebra $sl(n)$. Therefore one can conclude that, except for the usual MT  model, a multi-flavor generalization ($n_{F}\ge 2$, $n_{F}=$number of flavors) of the massive Thirring model allows certain zero-curvature formulations  only for its various constrained sub-models, in the both  NC and ordinary space-time descriptions. 

Except for the NCSG$_{1,2}$ models, which must correspond to the NCMT$_{1,2}$ models,  whose Lax pair formulations have already been provided in the literature, we have not been able to find the Lax pair formulations of the NCGSG$_{1,2}$ remaining sub-models. The relevant scalar field models, and their Lax pair formulations, which must be the counterparts of the (constrained) NC(c)GMT$_{1,2}$ and NC(c)BL$_{1,2}$ models are missing; if such Lax pairs exist they are expected to contain certain nonlocal expressions of the fields of the NCGSG$_{1,2}$ models. These points deserve a careful consideration in future research.

Various aspects of the models studied above deserve attention in
future research, e.g. the NC solitons and kinks of the NCATM$_{1,2}$
models and their relations with the confinement mechanism studied in ordinary space \cite{jhep4}, the bosonization of the NCGMT$_{1,2}$ and their sub-models, the NC zero-curvature formulation of the bosonic sector of the NC(c)GMT$_{1,2}$ and NC(c)BL$_{1,2}$ models,  as discussed above. In
particular, in the bosonization process of the NCGMT$_{1, 2}$ models, initiated in  \cite{jhep3} for the NCMT$_{1,2}$ case, we believe that
a careful understanding of the star-localized NC Noether
symmetries, as well as the classical soliton spectrum would be
desirable. In view of the rich spectra and relationships present in the above models it could be interesting to apply and improve some quantization methods, such as the one proposed in  \cite{guilarte}, in order to compute  the soliton and kink masses  quantum corrections. Another direction of research constitutes the NC zero-curvature formulations of the NCGMT$_{1,2}$ type models defined for Grassmannian fields.
\vskip .3in

{\sl Acknowledgments}

We thank G. Takacs and R. Schiappa for communications and valuable comments about their works, and  Prof. J. M. Guilarte for useful discussions about GSG models. HLC thanks IF-USP (S\~ao Paulo) for hospitality. HLC has been supported by FAPESP in the initial stage of the work and HB has been partially supported by CNPq.

\appendix

\section{GSG as a reduced affine Toda model coupled to matter}
\label{atm}

We provide the algebraic construction of the
 $sl(3, \IC)$  conformal affine Toda model coupled to matter fields (CATM) following refs. \cite{jhep1, matter}. The
reduction process to arrive at the classical GSG model closely  follows the ref. \cite{jhep4}. The $sl(3, \IC)$ CATM model is a two-dimensional field theory involving
four scalar fields and six Dirac spinors. The interactions among
the fields are as follows: 1) in the scalars equations of motion
there are the coupling of bilinears in the spinors to exponentials
of the scalars. 2) Some of the equations of motion for the spinors
have certain bilinear terms in the spinors themselves. That fact
makes it difficult to find a local Lagrangian for the theory.
Nevertheless, the model presents a lot of symmetries. It is
conformally invariant, possesses local gauge symmetries as well as
vector and axial conserved currents bilinear in the spinors. One
of the most remarkable properties of the model is that it presents
an equivalence between a U(1) vector conserved current, bilinear
in the spinors, and a topological currents depending only on the
first derivative of some scalars. This property allow us to
implement a bag model like confinement mechanism resembling what
one expects to happen in QCD. The model possesses a zero-curvature
representation based on the $\hat{sl}3(C)$ affine Kac� Moody
algebra. It constitutes a particular example of the so-called
conformal affine Toda models coupled to matter fields which has
been introduced in \cite{matter}. The corresponding model
associated to $\hat{sl}2(C)$ has been studied in \cite{nucl1}
where it was shown, using bosonization techniques, that the
equivalence between the currents holds true at the quantum level
and so the confinement mechanism does take place in the quantum
theory.

The off-critical affine Toda model coupled to matter (ATM) is
defined by gauge fixing the conformal symmetry \cite{annals} and imposing certain constraints in order to write a local Lagrangian for the model \cite{jhep1}. These treatments
of the $sl(3, \IC)$ ATM model used the symplectic and on-shell
decoupling methods to unravel the classical generalized
sine-Gordon  (GSG) and generalized massive Thirring (GMT) dual
theories describing the strong/weak coupling sectors of the ATM
model \cite{jmp, jhep1, annals}. As mentioned above the ATM model
describes some scalars coupled to spinor (Dirac) fields in which
the system of
 equations of motion has a local
gauge symmetry. Conveniently gauge fixing the local symmetry by
setting some spinor bilinears to constants we are able to decouple
the scalar (Toda) fields from the spinors, the final result is a
direct construction of the classical generalized sine-Gordon model
(GSG) involving only the scalar fields. In the spinor sector we
are left with a system of equations in which the Dirac fields
couple to the GSG fields. Another instance in which the quantum
version of the generalized sine-Gordon theory arises is in the
process of bosonization of the generalized massive Thirring model
(GMT), which is a multi-flavor extension of the usual massive
Thirring model such that, apart from the usual current-current
self-interaction for each flavor, it presents current-current
interactions terms among the various U(1) flavor currents
\cite{epjc}.

The zero-curvature condition (\ref{zeroc}) supplied with the potentials (\ref{aa1}) 
 gives the following equations of motion for the CATM model
\cite{matter} \br \label{eqnm1} \frac{\partial
^{2}\phi_{a}}{4i\,e^{\eta}} &=&m_{1}[e^{\eta
-i\th_{a}}\widetilde{\psi }_{R}^{l}\psi
_{L}^{l}+e^{i\th_{a}}\widetilde{\psi }_{L}^{l}\psi
_{R}^{l}]+m_{3}[e^{-i\th_{3}}\widetilde{\psi } _{R}^{3}\psi
_{L}^{3}+e^{\eta +i\th_{3}}\widetilde{\psi } _{L}^{3}\psi
_{R}^{3}];\,\,\,\,a=1,2\,\,\,\,\,\,\,\,\,\,\,\,\,\,\,\,\,\,\,
\\
\label{eqnm3} -\frac{\partial ^{2}\widetilde{\nu }}{4}
&=&im_{1}e^{2\eta -\th_{1}}\widetilde{\psi }_{R}^{1}\psi
_{L}^{1}+im_{2}e^{2\eta -\th_{2}}\widetilde{\psi }_{R}^{2}\psi
_{L}^{2}+im_{3}e^{\eta -\th_{3}}\widetilde{\psi } _{R}^{3}\psi
_{L}^{3}+{\bf m}^{2} e^{3\eta },\,\,
\\
\label{eqnm4} -2\partial _{+}\psi _{L}^{1}&=&m_{1} e^{\eta
+i\th_{1}}\psi _{R}^{1},\,\,\,\,\,\,\,\,\,\,\,\,\,\,\, -2\partial
_{+}\psi _{L}^{2}\,=\,m_{2}e^{\eta +i\th_{2}}\psi _{R}^{2},
\\
\label{eqnm5} 2\partial _{-}\psi _{R}^{1}&=&m_{1} e^{2\eta
-i\th_{1}}\psi _{L}^{1}+2i \(\frac{m_{2} m_{3}}{i
m_{1}}\)^{1/2}e^{\eta }(-\psi _{R}^{3} \widetilde{\psi
}_{L}^{2}e^{i\th _{2}}- \widetilde{\psi }_{R}^{2}\psi
_{L}^{3}e^{-i\th_{3}}),
\\
\label{eqnm7} 2\partial _{-}\psi _{R}^{2}&=&m_{2} e^{2\eta
-i\th_{2}}\psi _{L}^{2}+2i\(\frac{m_{1}
m_{3}}{im_{2}}\)^{1/2}e^{\eta }(\psi _{R}^{3} \widetilde{\psi
}_{L}^{1}e^{i\th_{1}}+ \widetilde{\psi }_{R}^{1}\psi
_{L}^{3}e^{-i\th_{3}}),
\\
\label{eqnm8} -2\partial _{+}\psi_{L}^{3}&=&m_{3} e^{2\eta +i\th
_{3}}\psi _{R}^{3}+2i\(\frac{m_{1} m_{2}}{im_{3}}\)^{1/2}e^{\eta
}(-\psi _{L}^{1}\psi _{R}^{2}e^{i\th_{2}}+\psi _{L}^{2}\psi
_{R}^{1}e^{i\th_{1}}),
\\
\label{eqnm9} 2\partial _{-}\psi _{R}^{3}&=&m_{3}e^{\eta
-i\th_{3}}\psi _{L}^{3},\,\,\,\,\,\,\,\,\,\,\,\, 2\partial
_{-}\widetilde{\psi }_{R}^{1}\,=\,m_{1} e^{\eta
+i\th_{1}}\widetilde{\psi }_{L}^{1},
\\
\label{eqnm10} -2\partial _{+}\widetilde{\psi }_{L}^{1} &=&m_{1}
e^{2\eta -i\th_{1}}\widetilde{\psi }_{R}^{1}+2i\(\frac{m_{2}
m_{3}}{i m_{1} }\)^{1/2}e^{\eta }(-\psi _{L}^{2}\widetilde{\psi
}_{R}^{3}e^{-i\th_{3}}-\widetilde{\psi }_{L}^{3}\psi
_{R}^{2}e^{i\th_{2}}),
\\
\label{eqnm12} -2\partial _{+}\widetilde{\psi }_{L}^{2}
&=&m_{2}e^{2\eta -i\th_{2}}\widetilde{\psi
}_{R}^{2}+2i\(\frac{m_{1} m_{3}}{i m_{2}} \)^{1/2}e^{\eta }(\psi
_{L}^{1}\widetilde{\psi }_{R}^{3}e^{-i\th_{3}}+\widetilde{\psi
}_{L}^{3}\psi _{R}^{1}e^{i\th_{1}}),
\\
\label{eqnm13} 2\partial _{-}\widetilde{\psi }_{R}^{2}&=&m_{2}
e^{\eta+i\th_{2}}\widetilde{\psi }_{L}^{2},
\,\,\,\,\,\,\,\,\,\,\,\,\,\,\,\, -2\partial _{+}\widetilde{\psi
}_{L}^{3}\,=\,m_{3} e^{\eta -i\th_{3}}\widetilde{\psi }_{R}^{3},
\\
\label{eqnm15} 2\partial _{-}\widetilde{\psi }_{R}^{3} &=&m_{3}
e^{2\eta +i\th_{3}}\widetilde{\psi
}_{L}^{3}+2i\(\frac{m_{1}m_{2}}{im_{3}} \)^{1/2}e^{\eta
}(\widetilde{\psi} _{R}^{1}\widetilde{\psi
}_{L}^{2}e^{i\th_{2}}-\widetilde{\psi }_{R}^{2}\widetilde{\psi
}_{L}^{1}e^{i\th_{1}}),
\\
\label{eqnm16}
\partial^{2}\eta&=&0,
\er where $\th_{1}\equiv2 \phi_{1}-\phi_{2},\,\th_{2}\equiv
2\phi_{2}-\phi_{1},\,\th_{3} \equiv \phi_{1}+\phi_{2}$. Therefore,
one has \br \th_{3}=\th_{1}+\th_{2}\label{phi123}\er

The $\phi$ fields are considered to be in general complex
fields. In order to define the classical generalized sine-Gordon
model we will consider these fields to be real.

Apart from the {\sl conformal invariance} the above equations
exhibit the $\(U(1)_{L}\)^{2}\otimes \(U(1)_{R}\)^{2}$ {\sl
left-right local gauge symmetry} \br \label{leri1}
\phi_{a} &\ra& \phi_{a} + \xi_{+}^{a}( x_{+}) + \xi_{-}^{a}( x_{-}),\,\,\,\,a=1,2\\
\widetilde{\nu} &\ra& \widetilde{\nu}\; ; \qquad \eta \ra \eta \\
\psi^{i} &\ra & e^{i( 1+ \gamma_5) \Xi_{+}^{i}( x_{+})
+ i( 1- \gamma_5) \Xi_{-}^{i}( x_{-})}\, \psi^{i},\label{leri3}\\
\,\,\,\, \widetilde{\psi}^{i} &\ra& e^{-i( 1+ \gamma_5)
(\Xi_{+}^{i})( x_{+})-i ( 1- \gamma_5) (\Xi_{-}^{i})(
x_{-})}\,\widetilde{\psi}^{i},\,\,\, i=1,2,3;\label{leri4}
\\
&&\Xi^{1}_{\pm}\equiv \pm \xi_{\pm}^{2} \mp
2\xi_{\pm}^{1},\,\,\Xi^{2}_{\pm}\equiv \pm \xi_{\pm}^{1}\mp
2\xi_{\pm}^{2},\,\,\Xi_{\pm}^{3}\equiv
\Xi_{\pm}^{1}+\Xi_{\pm}^{2}. \nonumber\er

One can get global symmetries for $\xi_{\pm}^{a}=\mp
\xi_{\mp}^{a}=$ constants. For a model defined by a Lagrangian
these would imply the presence of two vector and two chiral
conserved currents. However, it was found only half of such
currents \cite{bueno}. This is a consequence of the lack of a
Lagrangian description for the $sl(3)^{(1)}$ CATM in terms of the
$B$ and $F^{\pm}$ fields (however see Appendix \ref{localatm} for a local Lagrangian description of an off-critical and constrained sub-model). So, the vector current
\br \label{vec} J^{\mu}= \sum_{j=1}^{3} m_{j}
\bar{\psi}^{j}\gamma^{\mu}\psi^{j}\er and the chiral current \br
\label{chi} J^{5\,\mu} = \sum_{j=1}^{3} m_{j}
\bar{\psi}^{j}\gamma^{\mu}\gamma_{5} \psi^{j}+ 2 \partial_{\mu}
(m_{1}\phi_{1}+m_{2} \phi_{2})\er are conserved \br
\label{conservation} \pa_{\mu} J^{\mu}=0,\,\,\,\,\,\pa_{\mu}
J^{5\, \mu}=0.\er

The conformal symmetry is gauge fixed by setting \cite{annals} \br
\eta = \mbox{const}. \label{eta}\er

The off-critical ATM model obtained in this way exhibits the
vector and topological currents equivalence \cite{matter, annals}
\br \label{equivalence} \sum_{j=1}^{3} m_{j}
\bar{\psi}^{j}\gamma^{\mu}\psi^{j} \equiv \epsilon^{\mu
\nu}\partial_{\nu}
(m_{1}\phi_{1}+m_{2}\phi_{2}),\,\,\,\,\,\,\, m_{3}=m_{1}+
m_{2},\,\,\,\,m_{i}>0. \er

In the next steps we implement the reduction process to get the
GSG model through a gauge fixing of the ATM theory \cite{jhep4}. The local
symmetries (\ref{leri1})-(\ref{leri4}) can be gauge fixed through
\br \label{gf} i \bar{\psi}^{j}\psi^{j}= i
A_{j}=\mbox{const.};\,\,\,\,\,\,\bar{\psi}^{j}\gamma_{5}\psi^{j}
=0. \er

From the gauge fixing (\ref{gf}) one can write the following
bilinears
 \br
 \label{bilinears}
 \widetilde{\psi}_{R}^{j} \psi_{L}^{j} +
 \widetilde{\psi}_{L}^{j}\psi_{R}^{j}=0,\,\,\,\,\,j=1,2,3;
 \er
so,  the eqs. (\ref{gf})  effectively comprises three gauge fixing
 conditions.

It can be directly verified that the gauge fixing (\ref{gf})
preserves the currents conservation laws (\ref{conservation}),
i.e. from the equations of motion (\ref{eqnm1})-(\ref{eqnm16})
 and the gauge fixing (\ref{gf}) together with (\ref{eta}) it is possible
 to obtain the currents conservation laws (\ref{conservation}).

Taking into account the constraints (\ref{gf}) in the scalar
sector, eqs. (\ref{eqnm1}), we arrive at
 the following system
of equations (set $\eta=0$)\br \label{sys1} \pa^2 \phi_{1} &=&
M_{\psi}^{1}\,
\mbox{sin} (2\phi_{1}-\phi_{2}) + M_{\psi}^{3}\, \mbox{sin} (\phi_{1}+\phi_{2}),\\
\label{sys2}\,\,\,\pa^2 \phi_{2} &=& M^{2}_{\psi}\, \mbox{sin}
(2\phi_{2}-\phi_{1}) + M^{3}_{\psi}\, \mbox{sin}
(\phi_{1}+\phi_{2}),\,\,\,\, M^{i}_{\psi} \equiv 4 A_{i}\,
m_{i},\,\,\,\,i=1,2,3.\er

The system of equations above considered for real fields
$\phi_{1,\,2}$ as well as for real parameters $M_{\psi}^{i}$ defines the {\sl generalized sine-Gordon model} (GSG).

\section{The zero-curvature formulation of the $\hat{sl}(3)$ CATM model}
\label{atmapp}

We summarize the  zero-curvature formulation of the $\hat{sl}(3)$ CATM
model \cite{matter, bueno}. Consider the zero-curvature
condition \br \label{zeroc}
\partial_{+}A_{-}-\partial _{-}A_{+}+[A_{+},A_{-}]=0.
\er

The potentials take the form \br \label{aa1} A_{+}=-B
F^{+}B^{-1},\quad A_{-}=-\partial _{-}BB^{-1}+F^{-},\qquad \er
with \br \label{aa2} F^{+} \,=\,F_{1}^{+}+F_{2}^{+},\,\,\,\,\,\,
F^{-} \,=\, F_{1}^{-}+F_{2}^{-}, \er where $B$ and $F_{i}^{\pm }$
contain the fields of the model. Let us define
\br F^{\pm}_{m}&=&
\mp [E_{\pm 3}\,,\, W_{3-m}^{\mp}]\label{fw}\\
 E_{\pm 3}&=& \frac{1}{6} [(2 m_{1}+m_{2}) H^{\pm 1}_{1} +
(2m_{2}
+ m_{1}) H^{\pm 1}_{2} ],\,\,\,\,\,\,\,m_{3}=m_{1}+ m_{2} \label{ee1}\\
W^{-}_{1}&=& -\sqrt{\frac{4i}{m_{3}}} \psi^{3}_{R}
E^{-1}_{\alpha3} + \sqrt{\frac{4i}{m_{1}}}\widetilde{\psi}^{1}_{R}
E^{0}_{-\alpha1}+ \sqrt{\frac{4i}{m_{2}}}\widetilde{\psi}^{2}_{R}
E^{0}_{-\alpha2}
\label{ww1}\\
W^{+}_{1}&=& \sqrt{\frac{4i}{m_{1}}}\psi^{1}_{L} E^{0}_{\alpha1} +
\sqrt{\frac{4i}{m_{2}}}\psi^{2}_{L} E^{0}_{\alpha2}-
\sqrt{\frac{4i}{m_{3}}}\widetilde{\psi}^{3}_{L} E^{1}_{-\alpha3}
\label{ww2}\\
W^{-}_{2}&=& -\sqrt{\frac{4i}{m_{1}}} \psi^{1}_{R}
E^{-1}_{\alpha1} - \sqrt{\frac{4i}{m_{2}}}{\psi}^{2}_{R}
E^{-1}_{\alpha2}+ \sqrt{\frac{4i}{m_{3}}}\widetilde{\psi}^{3}_{R}
E^{0}_{-\alpha3}
\label{ww3}\\
W^{+}_{2}&=& \sqrt{\frac{4i}{m_{3}}}  \psi^{3}_{L} E^{0}_{\alpha3}
- \sqrt{\frac{4i}{m_{1}}}\widetilde{\psi}^{1}_{L}
E^{1}_{-\alpha1}- \sqrt{\frac{4i}{m_{2}}}\widetilde{\psi}^{2}_{L}
E^{1}_{-\alpha2} \label{ww4}\\
B &=& e^{i\th_{1} H^{0}_{1}+i\th_{2} H^{0}_{2}
}\,e^{\widetilde{\nu }C}\,e^{\eta
  Q_{ppal}}\,\,\equiv\,\, g\, e^{\widetilde{\nu }C}\,e^{\eta
  Q_{ppal}}. \label{equan1}
\er

$E_{\alpha _{i}}^{n},H^{n}_{1},H^{n}_{2}$ and  $C$ ($i=1,2,3; \,
n=0,\pm 1$) are some generators of $sl(3)^{(1)}$; $Q_{ppal}$ being
the principal gradation operator. The commutation relations for an
affine Lie algebra in the Chevalley basis are \br
&&\left[ \emph{H}_a^m,\emph{H}_b^n\right] =mC\frac{2}{\alpha_{a}^2}K_{a b}\delta _{m+n,0}  \label{a7}\\
&&\left[ \emph{H}_a^m,E_{\pm \alpha}^n\right] = \pm K_{\alpha
a}E_{\pm \alpha}^{m+n}
\label{a8}\\
&&\left[ E_\alpha ^m,E_{-\alpha }^n\right] =\sum_{a=1}^rl_a^\alpha
\emph{H}_a^{m+n}+\frac 2{\alpha ^2}mC\delta _{m+n,0}  \label{a9}
\\
&&\left[ E_\alpha ^m,E_\beta ^n\right] = \varepsilon (\alpha
,\beta )E_{\alpha +\beta }^{m+n};\qquad \mbox{if }\alpha +\beta
\mbox{ is a root \qquad }  \label{a10}
\\
&&\left[ D,E_\alpha ^n\right] =nE_\alpha ^n,\qquad \left[ D,\emph{H}%
_a^n\right] =n\emph{H}_a^n.  \label{a12} \er where $K_{\alpha
a}=2\a.\a_{a}/\a_{a}^2=n_{b}^{\a}K_{ba}$, with $n_{a}^{\a}$ and
$l_a^\alpha$ being the integers in the expansions
$\a=n_{a}^{\a}\a_{a}$ and $\a/\a^2=l_a^\alpha\a_{a}/\a_{a}^2$, and
$\varepsilon (\alpha ,\beta )$ the relevant structure constants.

Take $K_{11}=K_{22}=2$ and $K_{12}=K_{21}=-1$ as the Cartan matrix
elements of the simple Lie algebra $sl(3)$. Denoting by $\a_{1}$
and $\a_{2}$ the simple roots and the highest one by $\psi
(=\a_{1}+\a_{2})$, one has $l_{a}^{\psi}=1(a=1,2)$, and $K_{\psi
1}=K_{\psi 2}=1$. Take $\varepsilon (\alpha ,\beta )=-\varepsilon
(-\alpha ,-\beta ),\,\, \varepsilon_{1,2}\equiv \varepsilon
(\alpha_{1} ,\a_{2})=1,\,\, \varepsilon_{-1,3}\equiv
\varepsilon(-\alpha_{1} ,\psi )=1\,\, \mbox{and}\,
\,\,\varepsilon_{-2,3}\equiv \varepsilon (-\alpha_{2} ,\psi)=-1$.

One has $Q_{\mathrm{ppal}} \equiv \sum_{a=1}^{2}  {\bf
s}_{a}\l^{v}_{a}.H + 3 D$, where $\l^{v}_{a}$ are the fundamental
co-weights of $sl(3)$, and the principal gradation vector is ${\bf
s}=(1,1,1)$ \cite{kac}. This gradation decomposes
$\widehat{\mathfrak{sl}_3(\mathbb C)}$ into the following
subspaces \be \cgh_0 = \mathbb C \, H_1 \oplus \mathbb C \, H_2
\oplus \mathbb C \, C \oplus \mathbb C \, D = \mathbb C \, H_1
\oplus \mathbb C \, H_2 \oplus \mathbb C \, C \oplus \mathbb C \,
Q_{\mathrm{ppal}}, \label{zero} \ee and
\begin{eqnarray}
\cgh_{3m} &=& \mathbb C \, H_1^m \oplus \mathbb C \, H_2^m, \quad
m \ne 0,\label{grades1}
\\
\cgh_{3m+1} &=& \mathbb C \, E_{\a_1}^m \oplus \mathbb C \,
E_{\a_2}^m
\oplus \mathbb C \, E_{-\a_3}^{m+1}, \label{grades2}\\
\cgh_{3m+2} &=& \mathbb C \, E_{-\a_1}^{m+1} \oplus \mathbb C \,
E_{-\a_2}^{m+1} \oplus \mathbb C \, E_{\a_3}^{m}. \label{grades3}
\end{eqnarray}

\section{The off-critical and constrained $sl(3)$ ATM model}
\label{localatm}

The off-critical and constrained $sl(3)$ {\sl  affine Toda model coupled to
   matter fields} (ATM) is defined by the action \cite{jhep1}
\br \nonumber
\frac{1}{k} I_{\mbox{ATM}}^{(3)} &=& I_{WZNW}[g] + \int_{M}d^2x \{ \sum_{m=1}^{2} \Big[   <F_{m}^{-}\,,\, g F^{+}_{m} g^{-1}>\\
&& \nonumber
-\frac{1}{2} <E_{-3}\, , \, [W^{+}_{m}\, , \, \pa_{+} W^{+}_{3-m}]> + <F_{m}^{-}\, , \, \pa_{+} W^{+}_{m}> \\
&& +\frac{1}{2} <[W^{-}_{m}\,,\, \pa_{-} W^{-}_{3-m}]\,
 , \, E_{3}> + <\pa_{-} W^{-}_{m}
 \, , \, F_{m}^{+} > \Big]\},
\label{latm} \er where \br \label{wzw} I_{WZNW}[g]\,=\,
\frac{1}{8} \int_{M} d^2x Tr(\pa_{\mu} g \pa^{\mu} g^{-1}) +
\frac{1}{12} \int_{D} d^3 x\, \epsilon^{ijk} Tr(g^{-1} \pa_{i} g
 g^{-1} \pa_{j} g  g^{-1} \pa_{k} g), \er is the
Wess-Zumino-Novikov-Witten (WZNW) action for the matrix scalar
field of the model. The first term inside the summation of
(\ref{latm}) defines the form of the interactions  and the
remaining terms are the kinetic terms for the matrix fields
associated to the spinors. The equations of motion derived from
this action 
\br
\label{atm111}
 \pa_{-}(g^{-1} \pa_{+} g)& =& \sum_{m=1}^{2}\[F^{-}_{m}\,,\, g F^{+}_{m} g^{-1}\]
\\
\label{eqmf111}
\partial_{+} F^{-}_{m} &=& [E_{-3}, \partial_{+} W^{+}_{3-m}
],\,\,\,\,\,\,\,\,\,
\partial_{-} F^{+}_{m} = -[E_{3}, \partial_{-} W^{-}_{3-m} ],
 \\
 \partial_{+} W^{+}_{m} &=& - g  F^{+}_{m}  g^{-1},\,\,\,\,\,\,\,\,\,\,
  \partial_{-} W^{-}_{m} = - g^{-1}  F^{-}_{m}  g \label{eqmw111},
                \end{eqnarray}
are equivalent to the above CATM equations of motion (\ref{eqnm1})-(\ref{eqnm16})
 provided the following  constraints
\br \eta &=&0 \label{constcft}\\
 \[F_{2}^{\pm}\,,\,
g^{\mp 1}F_{1}^{\mp}g^{\pm 1}\]&=&0\label{constlocal},\er
are imposed. The first constraint defines an off-critical model, whereas the second ones allow a local Lagrangian description of the model.
Let us emphasize that the constraints (\ref{constlocal}) amount to drop all the terms with spinor bilinears on the right hand side of the set of equations (\ref{eqnm5})-(\ref{eqnm8}), (\ref{eqnm10})-(\ref{eqnm12}) and (\ref{eqnm15}), respectively. These constraints were introduced in refs. \cite{jmp, jhep1} since they are trivially satisfied by the soliton type  solutions of the full CATM model.    

\newpage

{\small
\begin{picture}(10224,4854)(300,-5500)
\put(500,-11400){\makebox(0,0)[lb]{\smash{\SetFigFont{11}{12.2}{it}
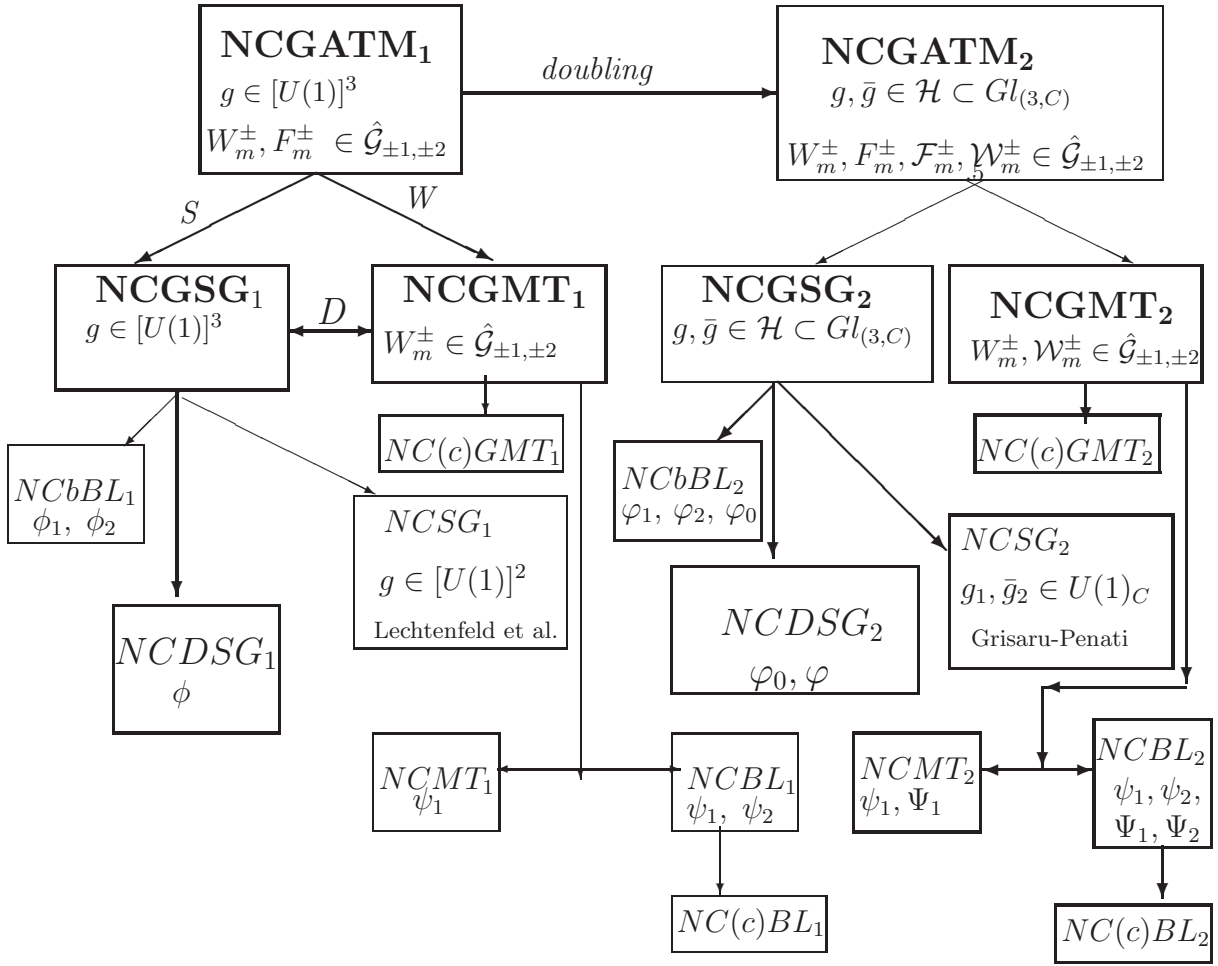
Fig. 1:  NCATM$_{1,2}$: dual sectors, sub-models and field contents.}}}\\ \\
\put(500,-12000){\makebox(0,0)[lb]{\smash{\SetFigFont{11}{12.2}{it}
Duality: {\bf S}={ strong sector}; {\bf
W}={ weak sector}; {\bf D}= { S - W
duality. A Lax pair is}}}}\\
\put(500,-12600){\makebox(0,0)[lb]{\smash{\SetFigFont{11}{12.2}{it}
available for NCSG$_{1,2}$/NCMT$_{1,2}$, NC(c)GMT$_{1,2}$ and  NC(c)BL$_{1,2}$, respectively.}}}
\put(500,-13200){\makebox(0,0)[lb]{\smash{\SetFigFont{11}{12.2}{it}
Dual sectors of the models NC(c)GMT$_{1,2}$,  NC(c)BL$_{1,2}$ and NCDSG$_{1,2}$ are missing }}}
\put(500,-13800){\makebox(0,0)[lb]{\smash{\SetFigFont{11}{12.2}{it} in the table above and deserve future
investigations.}}}

\put(3301,-3661){\makebox(0,0)[lb]{\smash{\SetFigFont{14}{16.8}{it}D}}}

\put(3880,-3961){\makebox(0,0)[lb]{\smash{\SetFigFont{12}{14.4}{it}
\,\,\,\,\,\,\,$W^{\pm}_{m}$}}}

\put(650,-4261){\makebox(0,0)[lb]{\smash{\SetFigFont{7}{14.4}{it}{}}}}
\put(5626,-1111){\makebox(0,0)[lb]{\smash{\SetFigFont{12}{14.4}{it}doubling}}}
\put(5626,-1600){\makebox(0,0)[lb]{\smash{\SetFigFont{12}{14.4}{it}}}}
\put(7000,-3811){\makebox(0,0)[lb]{\smash{\SetFigFont{12}{14.4}{it}$g,
\bar{g} \in {\cal H} \subset Gl_{(3,C)}$}}}
\put(6901,-4261){\framebox(2800,1200){}}

\put(8101,-2161){\framebox(4000,1800){}}
\put(8650,-1336){\makebox(0,0)[lb]{\smash{\SetFigFont{12}{14.4}{it}$g,
\bar{g} \in {\cal H} \subset Gl_{(3,C)}$}}}

\put(8200,-2010){\makebox(0,0)[lb]{\smash{\SetFigFont{12}{14.4}{}$W^{\pm}_{m},
F^{\pm}_{m}$, ${\cal F}^{\pm}_{m}$, ${\cal W}^{\pm}_{m}$}}}

\put(10051,-2161){\vector(2,-1){1740}}

\put(10051,-2161){\vector(-2,-1){1750}}

\put(900,-3800){\makebox(0,0)[lb]{\smash{\SetFigFont{11}{14.4}{}$g
 \in [U(1)]^{3}$}}}


 \put(1910,-4330){\vector(-1,-1){604}}
 \put(100,-5936){\framebox(1400,1005){}}
 \put(150,-5500){\makebox(0,0)[lb]{\smash{\SetFigFont{12}{14.4}{}{\small NCbBL}$_1$
}}}
  \put(350,-5805){\makebox(0,0)[lb]{\smash{\SetFigFont{12}{14.4}{}$\phi_1, \; \phi_2$
}}}

  \put(1910,-4430){\vector(2,-1){2004}}
 \put(3700,-7036){\framebox(2200,1575){}}
 \put(4000,-5826){\makebox(0,0)[lb]{\smash{\SetFigFont{12}{14.4}{}$NCSG_{1}
$}}}
\put(3950,-6426){\makebox(0,0)[lb]{\smash{\SetFigFont{12}{14.4}{}$g
\in [U(1)]^{2}$}}}

\put(3730,-6926){\makebox(0,0)[lb]{\smash{\SetFigFont{12}{14.4}{}\small{Lechtenfeld et al.}}}}

\put(2300,-1336){\makebox(0,0)[lb]{\smash{\SetFigFont{11}{14.4}{}$g
 \in [U(1)]^{3}$}}}

\put(3480,-1861){\makebox(0,0)[lb]{\smash{\SetFigFont{12}{14.4}{it}$\in
\hat{{\cal G}}_{\pm 1,\pm 2}$}}}
\put(2151,-1861){\makebox(0,0)[lb]{\smash{\SetFigFont{12}{14.4}{it}$W^{\pm}_{m},
F^{\pm}_{m}$}}}
\put(4201,-2461){\makebox(0,0)[lb]{\smash{\SetFigFont{12}{14.4}{it}W}}}
\put(7301,-3436){\makebox(0,0)[lb]{\smash{\SetFigFont{14}{16.8}{bf}
{\bf NCGSG}$_{2}$}}}
\put(4200,-3436){\makebox(0,0)[lb]{\smash{\SetFigFont{14}{16.8}{bf}{\bf NCGMT}$_{1}$}}}
\put(6051,-4261){\vector(0,-3){4140}}

 \put(6051,-8300){\vector(-2,0){870}}
 \put(3900,-8936){\framebox(1300,1005){}}
 \put(3950,-8500){\makebox(0,0)[lb]{\smash{\SetFigFont{12}{14.4}{}{\small NCMT}$_{1}$
}}}
  \put(4300,-8715){\makebox(0,0)[lb]{\smash{\SetFigFont{12}{14.4}{}$\psi_1 $
}}}

  \put(5060,-4196){\vector(0,-1){400}}
 \put(3960,-5196){\framebox(1900,575){}}
 \put(4000,-5086){\makebox(0,0)[lb]{\smash{\SetFigFont{12}{14.4}{}{\small NC(c)GMT}$_{1}
$}}}


 \put(6051,-8300){\vector(3,0){1040}}

 \put(7000,-8936){\framebox(1300,1005){}}
 \put(7150,-8500){\makebox(0,0)[lb]{\smash{\SetFigFont{12}{14.4}{}{\small NCBL}$_1$
}}}
  \put(7150,-8805){\makebox(0,0)[lb]{\smash{\SetFigFont{12}{14.4}{}$\psi_1,\; \psi_2$
}}}

  \put(7500,-8900){\vector(0,-1){700}}
 \put(7000,-10196){\framebox(1600,575){}}
 \put(7050,-9986){\makebox(0,0)[lb]{\smash{\SetFigFont{12}{14.4}{}{\small NC(c)BL}$_{1}
$}}}

\put(8551,-961){\makebox(0,0)[lb]{\smash{\SetFigFont{14}{16.8}{bf}{\bf NCATM}$_{2}$}}}
\put(2300,-886){\makebox(0,0)[lb]{\smash{\SetFigFont{14}{16.8}{bf}{\bf NCATM$_{1}}$}}} 
\thicklines 
\put(3001,-3736){\vector(-1,0){0}}
\put(3001,-3736){\vector(1,0){900}}
\put(4801,-1261){\vector(1,0){3300}}

\put(3301,-2086){\vector(2,-1){1830}}
\put(3301,-2086){\vector(-2,-1){1880}}

\put(2101,-2086){\framebox(2700,1725){}}

\put(3901,-4261){\framebox(2400,1200){}}
\put(601,-4336){\framebox(2400,1275){}}

\put(1000,-3436){\makebox(0,0)[lb]{\smash{\SetFigFont{14}{16.8}{bf}{\bf
 NCGSG$_{1}$}}}}

\put(1876,-2611){\makebox(0,0)[lb]{\smash{\SetFigFont{12}{14.4}{it}S}}}

\put(10100,-4036){\makebox(0,0)[lb]{\smash{\SetFigFont{11}{14.4}{it}$W^{\pm}_{m},
{\cal W}^{\pm}_{m}$}}}
\put(9901,-4261){\framebox(2600,1200){}}


\put(10300,-3600){\makebox(0,0)[lb]{\smash{\SetFigFont{14}{16.8}{bf}${\bf
N C G M T_{2}}$}}}

\put(12351,-4261){\vector(0,-1){3140}}
\put(12351,-7441){\vector(-1,0){1480}}
\put(10851,-7441){\vector(0,-1){840}}

 \put(11100,-8300){\vector(-1,0){870}}
 \put(8900,-8936){\framebox(1300,1005){}}
 \put(8950,-8400){\makebox(0,0)[lb]{\smash{\SetFigFont{12}{14.4}{}{\small NCMT}$_2$
}}}
  \put(8950,-8725){\makebox(0,0)[lb]{\smash{\SetFigFont{12}{14.4}{}$\psi_1, \Psi_1 $
}}}

 \put(11000,-8300){\vector(1,0){380}}
 \put(11400,-9096){\framebox(1200,1305){}}
 \put(11400,-8200){\makebox(0,0)[lb]{\smash{\SetFigFont{12}{14.4}{}{\small NCBL}$_2$
}}}
  \put(11590,-8605){\makebox(0,0)[lb]{\smash{\SetFigFont{12}{14.4}{}$\psi_1 ,\psi_2, $
}}}
\put(11590,-9005){\makebox(0,0)[lb]{\smash{\SetFigFont{12}{14.4}{}$\Psi_1
, \Psi_2 $ }}}

  \put(12100,-9096){\vector(0,-1){550}}
 \put(11000,-10296){\framebox(1600,575){}}
 \put(11050,-10086){\makebox(0,0)[lb]{\smash{\SetFigFont{12}{14.4}{}{\small NC(c)BL}$_{2}
$}}}

 \put(8050,-4300){\vector(-1,-1){580}}
 \put(6410,-5906){\framebox(1510,1005){}}
 \put(6470,-5380){\makebox(0,0)[lb]{\smash{\SetFigFont{12}{14.4}{}{\small NCbBL}$_2$
}}}
  \put(6470,-5680){\makebox(0,0)[lb]{\smash{\SetFigFont{12}{14.4}{}$\varphi_1 \, \varphi_2 \, \varphi_0$
}}}


\put(1851,-4360){\vector(0,-1){2140}}

\put(7500,-6900){\makebox(0,0)[lb]{\smash{\SetFigFont{14}{16.8}{bf}{\bf
{\small NCDSG}$_{2}$}}}}
 \put(7800,-7400){\makebox(0,0)[lb]{\smash{\SetFigFont{14}{16.8}{bf}{\bf
 $\vp_{0},\vp$ }}}}

\put(7001,-7500){\framebox(2550,1300){}}

\put(8051,-4261){\vector(0,-3){1840}}


 \put(1200,-7906){\framebox(1710,1305){}}
\put(1200,-7200){\makebox(0,0)[lb]{\smash{\SetFigFont{14}{16.8}{bf}{\bf
 {\small NCDSG}$_{1}$ }}}}

 \put(1800,-7600){\makebox(0,0)[lb]{\smash{\SetFigFont{12}{14.4}{}$\phi$
}}}

\put(8100,-4260){\vector(1,-1){1750}}

 \put(9900,-7236){\framebox(2300,1575){}}
 \put(10000,-6016){\makebox(0,0)[lb]{\smash{\SetFigFont{12}{14.4}{}$NCSG_{2}
$}}}
\put(10000,-6516){\makebox(0,0)[lb]{\smash{\SetFigFont{12}{14.4}{}$g_{1},
\bar{g}_{2} \in U(1)_{C}$}}}

\put(10000,-7016){\makebox(0,0)[lb]{\smash{\SetFigFont{12}{14.4}{}
{\small Grisaru-Penati}}}}

  \put(11300,-4296){\vector(0,-1){400}}
 \put(10160,-5196){\framebox(1900,575){}}
 \put(10160,-5086){\makebox(0,0)[lb]{\smash{\SetFigFont{12}{14.4}{}\small{NC(c)GMT}$_{2}
$}}}

\end{picture}
}

\end{document}